\documentclass[twocolumn,prb,eqsecnum,showpacs,floatfix]{revtex4}
\usepackage{amsmath,mathrsfs}
\usepackage{graphicx}
\begin{document}

\title{Effects of a nonlinear bath at low temperatures}

\author{Hanno Gassmann and C. Bruder}%
\affiliation{Department of Physics and Astronomy, University of Basel,
 Klingelbergstrasse 82, 4056 Basel, Switzerland}%

\date{version of November 2}

\begin{abstract}
We use the numerical flow-equation renormalization method to study a
nonlinear bath at low temperatures. The model of our nonlinear bath
consists of a single two-level system coupled to a linear oscillator
bath. The effects of this nonlinear bath are analyzed by coupling it
to a spin, whose relaxational dynamics under the action of the bath is
studied by calculating spin-spin correlation functions. As a first
result, we derive flow equations for a general four-level system
coupled to an oscillator bath, 
valid at low temperatures. We then treat
the two-level system coupled to our nonlinear bath as a special case
of the dissipative four-level system. We compare the effects of the
nonlinear bath with those obtained using an effective linear bath, and
study the differences between the two cases at low temperatures.
\end{abstract}



\maketitle

\section{Introduction}
The linear bath of harmonic oscillators is the most common model
system to describe a quantum environment or bath, leading to
dissipation and decoherence
\cite{Caldeira83b,Caldeira83c,Weiss00a}. In this paper we examine an
environment which is the simplest possible quantum-mechanical
non-oscillator bath \cite{Gassmann02}: a single two-level system
subject to an oscillator bath. The effects of this nonlinear bath are
analyzed by coupling it to a spin, whose relaxational dynamics under
the action of the bath is studied by calculating spin-spin correlation
functions.

Nonlinear baths have received a great deal of interest in the field of
quantum computation, e.g., in the context of superconducting devices
\cite{Paladino,makhlin03a} and for spin-based proposals
\cite{stamp,Khaetskii,coish04}.

There are two ways to define the border between system and bath
in our total system. The first is to consider one two-level system
to be the system and the other two-level system belonging to the
bath. The bath consists then of a two-level system and the oscillators
together and therefore it is a non-oscillator or nonlinear
bath. Studying the problem this way is relevant, since, in many
physical situations where the precise nature of the bath decohering a
given system is unknown, it is modeled as a linear bath, with some
given correlation function. It is, therefore, desirable to understand
in more detail the kind and magnitude of possible errors introduced by
such an approximation, in cases where the coupling cannot be assumed
to be weak.

The other way to define the system and bath is as follows: taken
together, the two-level systems form a four-level system that is
coupled to the linear oscillator bath. Such systems of two two-level
systems coupled to linear oscillator baths, representing a formidable
problem in itself, have been studied in-depth using different methods,
such as functional-integral approaches \cite{4}, master equations
\cite{5,5b}, and numerical calculations within the quasi-adiabatic
propagator path integral method \cite{6}. In this work we use a
different approach: the numerical flow-equation renormalization
method.

The flow-equation method itself was introduced by Wegner \cite{wegner}
and by Glasek and Wilson \cite{wilson}. Using the flow-equation
method one works in a Hamiltonian framework. The whole procedure is
non-perturbative, as it relies on a unitary transformation and has an
energy-scale separation built in. The extension of the flow-equation
formalism to dissipative systems was developed in
Refs.~\onlinecite{neu}, \onlinecite{kehrein}, and
\onlinecite{mielke}. The effective Hamiltonian simplifies due to the
fact that the coupling between bath and system disappears and
equilibrium functions can be calculated easily. The method is
non-perturbative, e.g., neither a Born nor Markoffian approximation
is applied, which is what makes it attractive and new. Some
drawbacks are that the flow equations for the observables are not
closed, and a special ansatz is needed. Here, we choose a linear ansatz,
which works only for low temperatures smaller than the typical
low-energy scale of the system Hamiltonian. What is new in our work is
that we apply the method to a general dissipative four-state system.
The flow equations are designed to yield
correlation functions for non-ohmic baths as well, which, as we will
see below, makes it possible to treat the linear bath approximation to
our nonlinear problem on the same footing. In
Refs.~\onlinecite{kehrmiel}, \onlinecite{Kehrein96c}, \onlinecite{stauber},
\onlinecite{kleffa}, and \onlinecite{kleffb}, subohmic, superohmic,
and a peaked structured environment, were studied with flow
equations. Similarly, the linear-bath approximation leads to a new
peaked structured bath spectral density, to our knowledge, never
studied before. The numerical integration of the flow equations
allows us to study the nonlinear bath model by analyzing the spin-spin
correlation functions. The numerical results are obtained at zero
temperature. These results are then compared with those obtained from
a linear bath with a non-ohmic bath spectral density. The
bath spectral density is chosen in such a way that the linear bath
leads to the same results as the original nonlinear bath in the limit
of small enough system-bath coupling strength. The bath and system
correlation functions are both calculated using flow equations. By
carrying out the comparison, we are able to study the effects of the
nonlinear bath at zero temperature.

The remainder of this work is organized as follows: First we describe
the model in Sec.~\ref{flow:sec:model}. Then, we write down the flow
equations for the general dissipative four-level system. The flow equations for the nonlinear-bath system
(Sec.~\ref{flow:sec:nonlinear}) and its linear approximation
(Sec.~\ref{flow:sec:linear}) are both special cases of these general
flow equations. Finally, we discuss the numerical details in
Sec.~\ref{flow:sec:numdet} and compare the results of the two
approaches in Sec.~\ref{flow:sec:complinnonlin}.

\section{The nonlinear bath model}\label{flow:sec:model}

As a model for a nonlinear bath we study a two-level system \(
S \) coupled to a bath, which 
consists of another two-level system
\( B \) coupling to a linear oscillator bath \( F \): 

\begin{eqnarray}\label{model}\nonumber
H&=&\epsilon _{S}\sigma ^{S}_{z}+\Delta
_{S}\sigma^{S}_{x}+g\sigma^{S}_{z}\sigma^{B}_{z}+\Delta \sigma^{B}_{x}
\\
&+&\sigma_{z}^{B}\sum_{k}\lambda_{k}(b_{k}+b_{k}^{\dagger})+
\sum_{k}\omega_{k}:b_{k}^{\dagger}b_{k}:\,.
\end{eqnarray}
Here the parameters \( \epsilon _{S} \) and \( \Delta _{S} \) serve to define
any desired two-level system \( S \). This system is coupled to \( B \) via
\( \sigma^{B}_{z} \), with the coupling strength between \( S \) and
\( B \) being given by \( g \). The coupling constants \(\lambda_{k}\)
are connected to the bath spectral density in the following way:
\begin{equation}
J(\omega)\equiv\pi\sum_{k}\lambda_{k}^2\delta(\omega-\omega_{k})=
\alpha\omega f(\omega/\omega_{c})\,,
\end{equation}
with the bath cutoff \(\omega_{c}\), the system-bath coupling strength
\(\alpha\) and the cutoff function \(f(x)\). The oscillations of \(
\sigma^{B}_{z}(t) \) at the frequency \( 2\Delta \) are noisy, due to
the action of the harmonic oscillator bath \(F\). The dissipative
dynamics of \( S \) can be characterized in terms of several different
quantities. Here we will analyze the decay of the equilibrium
correlator \(
\left\langle\sigma_{z}^{S}(t)\sigma_{z}^{S}(0)\right\rangle \) by
calculating its Fourier transform:
\begin{equation}
\label{Kzz}
K_{zz}^{S}(\omega)\equiv\frac{1}{2\pi }
\int _{-\infty }^{+\infty }dt\, e^{i\omega t}\left\langle
\sigma_{z}^{S}(t)\sigma_{z}^{S}(0)\right\rangle\,.
\end{equation}
\( K^{S}_{zz}(\omega )\) is real-valued, for zero temperature it
vanishes on the negative real axis and the integral over all
frequencies gives \( 1 \).

\section{Flow equations for the dissipative four-state system}
\label{flow:sec:nonlinear}
We will write down the flow equations for the most general four-level system
coupling to a linear oscillator bath.
Therefore many other systems can
be analyzed as special cases of this general framework. 
The Hamiltonian given in Eq.~(\ref{model}) is a special case of:
\begin{eqnarray}\label{genham}\nonumber
H=\sum_{\alpha \beta}\Delta^{\alpha \beta}\Sigma_{\alpha
\beta}+\sum_{k}\sum_{\alpha \beta}\lambda_{k}^{\alpha
\beta}\Sigma_{\alpha \beta}(b_{k}+b_{k}^{\dagger})
\\
+i\sum_{k}\sum_{\alpha \beta}\kappa_{k}^{\alpha \beta}
\Sigma_{\alpha \beta}(b_{k}-b_{k}^{\dagger})+\sum_{k}
\omega_{k}:b_{k}^{\dagger}b_{k}:\,.
\end{eqnarray}
The \(\Sigma_{\alpha \beta}\) are the tensor products of the Pauli
matrices \(\Sigma_{\alpha
\beta}=\sigma_{\alpha}\otimes\sigma_{\beta}\) and the Greek indices
are always summed from zero to three. \(\sigma_{0}\) is the unit
matrix. By using the flow-equation technique we approximately diagonalize the
model Hamiltonian given in Eq.~(\ref{genham}) by means of a
unitary transformation:
\begin{equation}
H(l)=U(l) H U(l)^{\dagger}\,.
\end{equation}
Here, \(l\) is the flow parameter which is roughly equivalent to
the square of the inverse
energy scale which is being decoupled. The unitary transformation can
be written in a differential form
\begin{equation}
\frac{d H(l)}{dl}=[\eta(l),H(l)]\,,
\end{equation}
where
\begin{equation}
\eta(l)=\frac{d U(l)}{dl}U^{-1}(l)\,.
\end{equation}
Using the anti-Hermitian generator \(\eta\) in the canonical form
\(\eta=[H_{0},H]\) leads to the fixed point
\(H(l\rightarrow\infty)=H_{0}\). 
We have investigated two different generators \(\widetilde{\eta}\) 
and \(\eta\), leading to two different fixed points, viz.,
\begin{equation}
\widetilde{H}_{0}=\sum_{k}\omega_{k}:b_{k}^{\dagger}b_{k}:
\end{equation}
the oscillator bath, 
and
\begin{equation}
H_{0}=\sum_{\alpha \beta}\Delta^{\alpha \beta}\Sigma_{\alpha
\beta}+\sum_{k}\omega_{k}:b_{k}^{\dagger}b_{k}:
\end{equation}
the total Hamiltonian without interaction terms. Note that we only
have to derive flow equations with the generator \(\eta\). The flow
equations using \(\widetilde{\eta}\) are contained in the latter, due
to the similar structure of the generators.

The commutator \([\eta,H]\) contains
coupling terms which are bilinear in the bosonic operators. We neglect
these terms by truncating the Hamiltonian after linear bosonic
terms. The bilinear terms are included by modifying the generator with
an additional bilinear term as follows \cite{stauber,kehrein}:
\begin{equation}
\eta=[H_{0},H]+\sum_{kq}\eta_{kq}
:(b_{k}+b_{k}^{\dagger})(b_{q}-b_{q}^{\dagger}):\,.
\end{equation}
The parameters \(\eta_{k q}\) are chosen in such a way that the
bilinear terms are not generated. Because this cannot be done exactly
we neglect terms which have the normal-ordered form: system operator
times a bilinear bosonic operator. Here the normal order is defined
with respect to the non-interacting Hamiltonian.

Comparing the general Hamiltonian given in Eq.~(\ref{genham}) with
Eq.~(\ref{model}) specifies the initial conditions for the flow
equations listed in Appendices \ref{flow:sec:flow-ham} and
\ref{eta}. The flow equations are nonlinear ordinary coupled
differential equations and they have the same structure as those
obtained in Ref.~\onlinecite{stauber} due to the similar truncation
procedure. The renormalization of the bath modes \(\omega_{k}\)
vanishes in the thermodynamic limit and can be neglected \cite{neu}.

In order to calculate correlation functions the observables
have to be subjected to the same sequence of infinitesimal
transformations as the Hamiltonian. The observable flow cannot be
closed and therefore a linear ansatz has to be chosen which is only
valid for temperatures smaller than the typical low-energy scale of
the four-level system coupling to the linear oscillator bath.
The most general linear ansatz which is possible and used in the
following is given by:
\begin{eqnarray}\nonumber
O=\sum_{\alpha \beta}h^{\alpha \beta}\Sigma_{\alpha
\beta}+\sum_{k}\sum_{\alpha \beta}\mu_{k}^{\alpha \beta}\Sigma_{\alpha
\beta}(b_{k}+b_{k}^{\dagger})
\\
+i\sum_{k}\sum_{\alpha \beta}\nu_{k}^{\alpha \beta}
\Sigma_{\alpha \beta}(b_{k}-b_{k}^{\dagger})\,.
\end{eqnarray}
The flow equations for the observables
\begin{equation}
\frac{d O(l)}{dl}=[\eta(l),O(l)]
\end{equation}
are closed according to the same normal ordering scheme as above and are
 listed in the Appendix \ref{flow:sec:flowO}. 
The correlation functions for \(l\rightarrow\infty\) are defined by:
\begin{equation}
\left\langle O(t)O\right\rangle=\frac{\mathrm{tr}\{e^{-\beta
H_{0}}e^{i t H_{0}}O e^{-i t H_{0}}O\}}{\mathrm{tr}\{e^{-\beta
H_{0}}\}}\,.
\end{equation}
With the non-interacting fixed-point Hamiltonian \(H_{0}\) such
correlation functions can be calculated easily.

Below we present the comparatively short result for the correlation
functions for the fixed point \(\widetilde{H}_{0}\). They are found to
be:
\begin{eqnarray}\nonumber
\left\langle O(t)O\right\rangle=h^{0 0}h^{0 0}+h^{0 i}h^{0 i}+h^{i
0}h^{i 0}+h^{i j}h^{i j}
\\
+\sum_{k}\Big[(n_{k}+1)e^{-i\omega_{k}t}c_{k}
+n_{k}e^{i\omega_{k}t}c_{k}\Big]\,,
\end{eqnarray}
where
\begin{eqnarray}\nonumber
c_{k}&=&\mu_{k}^{0
0}\mu_{k}^{0 0}
+\mu_{k}^{0 i}\mu_{k}^{0 i}+\mu_{k}^{i
0}\mu_{k}^{i 0}
+\mu_{k}^{i j}\mu_{k}^{i j}
\\\nonumber
&+&\nu_{k}^{0
0}\mu_{k}^{0 0}
+\nu_{k}^{0 i}\mu_{k}^{0 i}+\nu_{k}^{i
0}\mu_{k}^{i 0}
+\nu_{k}^{i j}\mu_{k}^{i j}
\\
&-&\mu_{k}^{0
0}\nu_{k}^{0 0}
-\mu_{k}^{0 i}\nu_{k}^{0 i}-\mu_{k}^{i
0}\nu_{k}^{i 0}
-\mu_{k}^{i j}\nu_{k}^{i j}
\\\nonumber
&-&\nu_{k}^{0
0}\nu_{k}^{0 0}
-\nu_{k}^{0 i}\nu_{k}^{0 i}-\nu_{k}^{i
0}\nu_{k}^{i 0}
-\nu_{k}^{i j}\nu_{k}^{i j}\,.
\end{eqnarray}
The Fourier transform of the correlation function is given by
\begin{eqnarray}\nonumber
&&\left\langle O O\right\rangle_{\omega}=\big(
h^{0 0}h^{0 0}+h^{0 i}h^{0 i}+h^{i
0}h^{i 0}+h^{i j}h^{i j}\big)\delta(\omega)
\\
&+&\sum_{k}\Big[(n_{k}+1)\delta(\omega-\omega_{k})c_{k}
+n_{k}\delta(\omega+\omega_{k})c_{k}\Big]\,,
\end{eqnarray}
where the Fourier transform is defined by
\begin{equation}
\left\langle O
O\right\rangle_{\omega}\equiv\frac{1}{2\pi}\int_{-\infty}^{\infty}dt e^{i\omega
t}\left\langle O(t)O\right\rangle\,.
\end{equation}
With these formulas different correlation functions can be calculated.
The above formulas can also be used to obtain correlation functions
for different operators, i.e., \(\left\langle O^{(1)}(t)O^{(2)}\right\rangle\).
The only difference in the result for \(c_{k}\) is that now a sum of
products of the form \(\mu_{k}^{(1) 0 i}\mu_{k}^{(2) 0 i}\)
occurs and therefore the number of flow equations is increased.
Below we put \(O=\sigma_{z}^{S}\), i.e., analyze the decay of the
equilibrium correlator 
\( \left\langle\sigma_{z}^{S}(t)\sigma_{z}^{S}(0)\right\rangle \). The
correlation function is determined by the initial conditions of the
flow equations. The limit of zero temperature \(T=0\) can be obtained
by putting \(n_{k}=0\) in the flow equations.

\section{The linear-bath approximation}\label{flow:sec:linear}
In this section we study an approximation to the nonlinear bath.
We do this by replacing the composite bath \(B\) and \(F\) by a
linear oscillator bath acting on the system \(S\). What we obtain is a
spin-boson system with a non-ohmic peaked bath spectral density
\begin{equation}
H=\Delta_{S} \sigma^{S}_{x}+\epsilon_{S}\sigma_{z}^{S}
+\sigma_{z}^{S}\sum_{k}\widetilde{\lambda}_{k}(b_{k}+b_{k}^{\dagger})+
\sum_{k}\omega_{k}:b_{k}^{\dagger}b_{k}:\,,
\end{equation}
where the bath spectral density is given by 
\begin{equation}
\widetilde{J}(\omega)\equiv\pi\left\langle B
B\right\rangle_{\omega}\equiv\pi\sum_{k}
\widetilde{\lambda}_{k}^2\delta(\omega-\omega_{k})\,.
\end{equation}
The linear-bath approximation should become exact in the weak-coupling
limit \(g\ll 1\) according to Ref.~\onlinecite{Caldeira93}. For higher
coupling \(g\) we expect deviations to the nonlinear bath even in the
thermodynamic limit \(N\rightarrow\infty\) due to, e.g., the
non-separable structure~\cite{makri99} of our nonlinear bath model. 

We start our discussion by calculating the bath correlation
function. The Fourier transform of the correlator of \(B\equiv
g\sigma^B_z\) defines the ``bath spectrum'' \(\left\langle B
B\right\rangle_{\omega}\).
For zero temperature, \(\left\langle B
B\right\rangle_{\omega}\) vanishes for negative frequencies.
The Hamiltonian of the bath alone is given by
\begin{equation}
H_{B F}=\Delta \sigma^{B}_{x}
+\sigma_{z}^{B}\sum_{k}\lambda_{k}(b_{k}+b_{k}^{\dagger})+
\sum_{k}\omega_{k}:b_{k}^{\dagger}b_{k}:\,.
\end{equation}
This is the initial Hamiltonian for the flow equations. For the
purpose of calculating the bath correlation function, one has to solve
the symmetric spin-boson problem. This cannot be done in a closed
analytical way \cite{leggett87,Weiss00a,grifoni99}. An alternative
method involves flow equations, which we apply again. We use the flow
equation system used in Ref.~\onlinecite{stauber} with a generator of
the form \(\widetilde{\eta}\).  The truncation scheme includes (as
before) all coupling terms which are linear in the bosonic modes. The
flow equation method is not restricted to any particular bath type,
i.e., the bath spectral density of the linear bath can be chosen
freely.

\begin{figure}
\includegraphics[width=80mm]{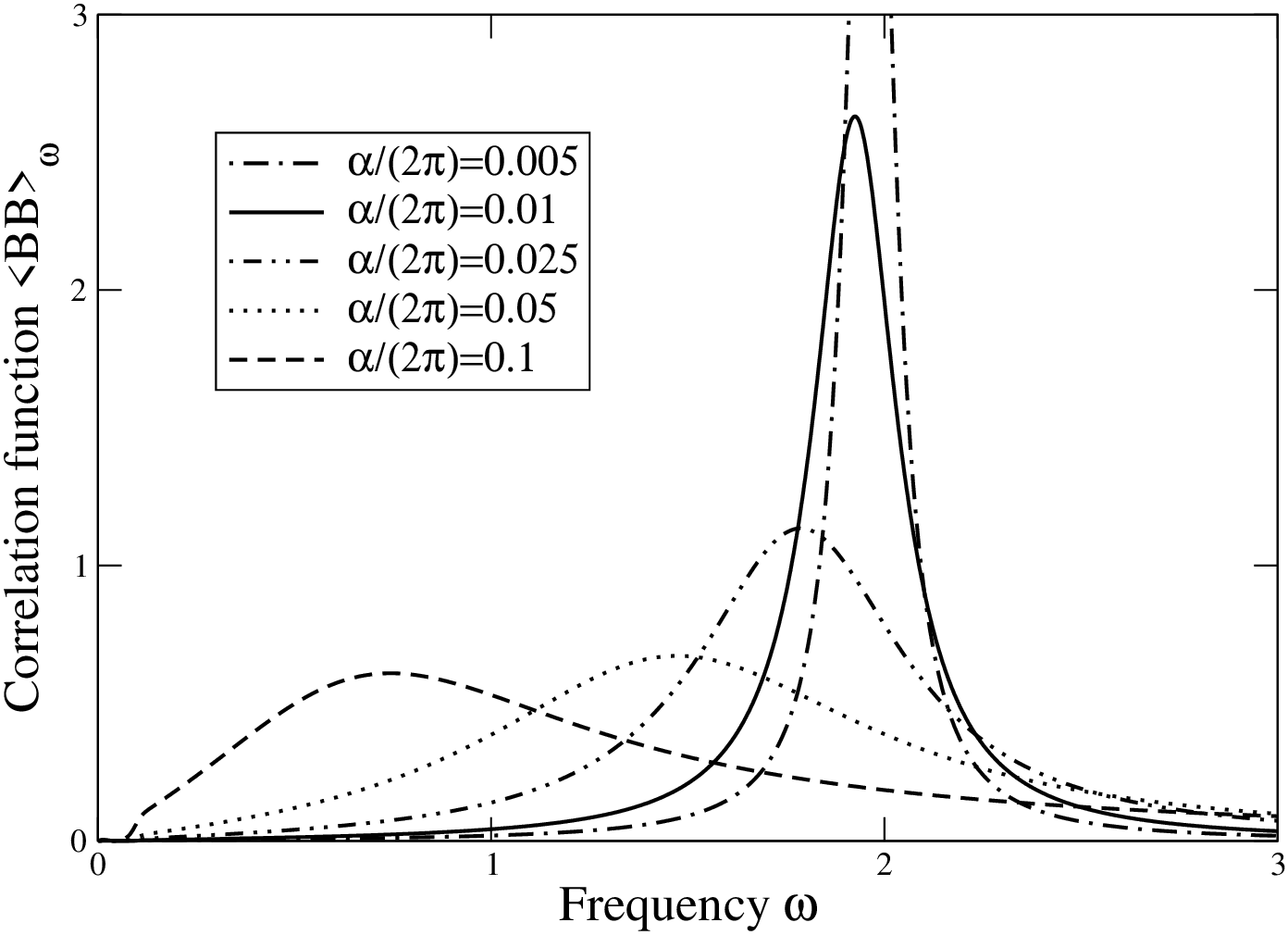}
\caption{\label{fig:bbcorrb}The \(\left\langle BB \right\rangle_{\omega
}\) correlation function for different values of \(\alpha\). The
parameters are: \(g=1\) and \(\Delta=1\). The
height of the \(\alpha=0.005\) peak is \(\sim 5.2\). }
\end{figure}
The results for the bath correlation function \(\left\langle B
B\right\rangle_{\omega}\) are shown in Fig.~\ref{fig:bbcorrb}.
\(\left\langle B B\right\rangle_{\omega}\) consists of one peak at the
frequency \(\approx 2\Delta\), for small coupling \(\alpha\). For
larger coupling it is shifted towards lower frequencies. Because 
the bath \(F\) is ohmic, the position of the maximum of
\(\left\langle B B\right\rangle_{\omega}\) is given approximately by
\(2\Delta (2\Delta/\omega_{c})^{2\alpha/(\pi-2\alpha)}\), which is a
result obtained within an adiabatic renormalization scheme
\cite{Weiss00a}. In our results, the relation is best fulfilled for
\(\alpha/(2\pi)\lesssim 0.025\). 
The low-frequency behavior predicted by the exact zero-temperature
Shiba relation can only be found by solving the
asymptotic flow equations \cite{stauber}, which is not done in this
work (note the deviation from linearity for small \(\omega\ll 1\) in
Fig.~\ref{fig:bbcorrb}, e.g., graph for \(\alpha/(2\pi)=0.1\). Thus, the bath spectral density shows a
resonance at a characteristic frequency and behaves ohmic at small
frequencies. This is why we expect the results to be comparable to
the ones obtained in Ref.~\onlinecite{kleffb,thorwart03}. But we
stress that here, in contrast, the linear bath is modified with a
second two-state system, compared to an additional oscillator.

To obtain the final result for the linear-bath approximation we use
the \(\left\langle B B\right\rangle_{\omega}\) correlation function as
bath spectrum \(\widetilde{J}(\omega)=\pi\left\langle B
B\right\rangle_{\omega}\) and solve the flow equations for the
spin-boson system with this non-ohmic spectrum. The results are
discussed in Section~\ref{flow:sec:complinnonlin}.

To close this section, we make a short note regarding higher-order
correlation functions with respect to the flow equations used
here. 
Higher-order correlation functions are important to understand
the behavior of a nonlinear bath. For small enough system-bath
coupling strength, the nonlinear bath is identical to a linear one
\cite{Caldeira93}. In this limit, the two-time bath correlation
function plays a crucial role. For higher coupling strength,
higher-order correlation functions become relevant. The linear bath
has the special feature that higher-order correlation functions
factorize into two-time correlation functions, since due to the
Gaussian nature of the linear bath, the cumulant expansion breaks off
at second order. The nonlinear case has generally no such feature,
but we will see that the correlation functions obtained with the flow
equations factorize. The reason for this is the choice of a linear
ansatz for the observable flow:
\begin{equation}
\sigma_{z}(l)=h^{03}(l)\sigma_{z}+\mu_{k}^{01}(l)\sigma_{x}
(b_{k}+b_{k}^{\dagger})\,.
\end{equation}
In the limit \(l\rightarrow\infty\) we find for the symmetric
spin-boson system
\begin{equation}
\sigma_{z}(l\rightarrow\infty)=\mu_{k}^{01}
(l\rightarrow\infty)\sigma_{x}(b_{k}+b_{k}^{\dagger})\,,
\end{equation}
and
\(H(l\rightarrow\infty)=\widetilde{H}_{0}=\sum_{k}\omega_{k}b_{k}^{\dagger}
b_{k}\).
For \(l\rightarrow\infty\) the observables behave like linear
oscillator variables and all higher-order correlation functions
factorize. 

\section{Details of the numerical calculation}\label{flow:sec:numdet}
The flow equations are coupled ordinary differential equations. The
integration method which turns out to be the most appropriate one is
the fourth-order Runge-Kutta method with variable step size
\cite{nrecipes}. The reason for this is that there are regions where
a certain subset of the differential equations does not change much
anymore and larger integration steps can be used, see discussion
below.

We have used a linear energy spacing \(\delta\omega\) for the bath
modes \(\omega_{k}\) and a sharp cutoff function
\(f(x)=\Theta(1-x)\). The results do not depend on these choices. The
bath modes \(\omega_{k}\) are then given by
\(\omega_{k}=(k-1/2)\delta\omega\) and the coupling constants
\(\lambda_{k}\) are found to be
\(\lambda_{k}=\big(\sqrt{\alpha/\pi}\sqrt{k-1/2}\big)\delta\omega\),
where \(\delta\omega=\omega_{c}/N\) and \(N\) is the number of bath
modes used for the numerics. For the four-level system the number of
differential equations is \(64 N+32\) in the most general case,
\(5N+7\) for the nonlinear bath with \(\epsilon_{S}=0\) (see Appendix
\ref{flow:sec:flowbias}), \(10N+14\) for the nonlinear bath with
\(\epsilon_{S}\neq 0\) and \(3 N+2\) for the symmetric spin boson
system. We have used up to \(N=10^4\) bath oscillators. For the
particular choice \(\epsilon_{S}=0\), the correlation function at
discrete frequencies is given by
\(K_{zz}^{S}(\omega_{k})=(\mu_{k}^{11})^2/\delta\omega\), for the
generator \(\widetilde{\eta}\). All the \(h^{\alpha \beta}\)
coefficients vanish for increasing flow \(l\rightarrow\infty\), as
expected. For the other generator \(\eta\), we have to evaluate
\begin{eqnarray}\nonumber
&K_{zz}^{S}(\omega)&=\sum_{n}|\langle
0|(h^{03}\Sigma_{03}+h^{13}\Sigma_{13})|n\rangle|^2\delta(\omega+E_{0}-E_{n})
\\
&+&\sum_{k,n}(\mu_{k}^{11})^2|\langle
0|\Sigma_{11}|n\rangle|^2\delta(\omega+E_{0}-E_{n}-\omega_{k})\,,
\end{eqnarray}
where \(E_{n}\) are the eigenvalues of \(H_{S}\) and \(|n\rangle\) are
the eigenvectors of \(H_{S}\) for \(l\rightarrow\infty\). In the
parameter regime considered below, it turns out that the
equilibrium correlator is given by the same expression as for the
generator \(\widetilde{\eta}\) with high accuracy.

\begin{figure}
\includegraphics[width=80mm]{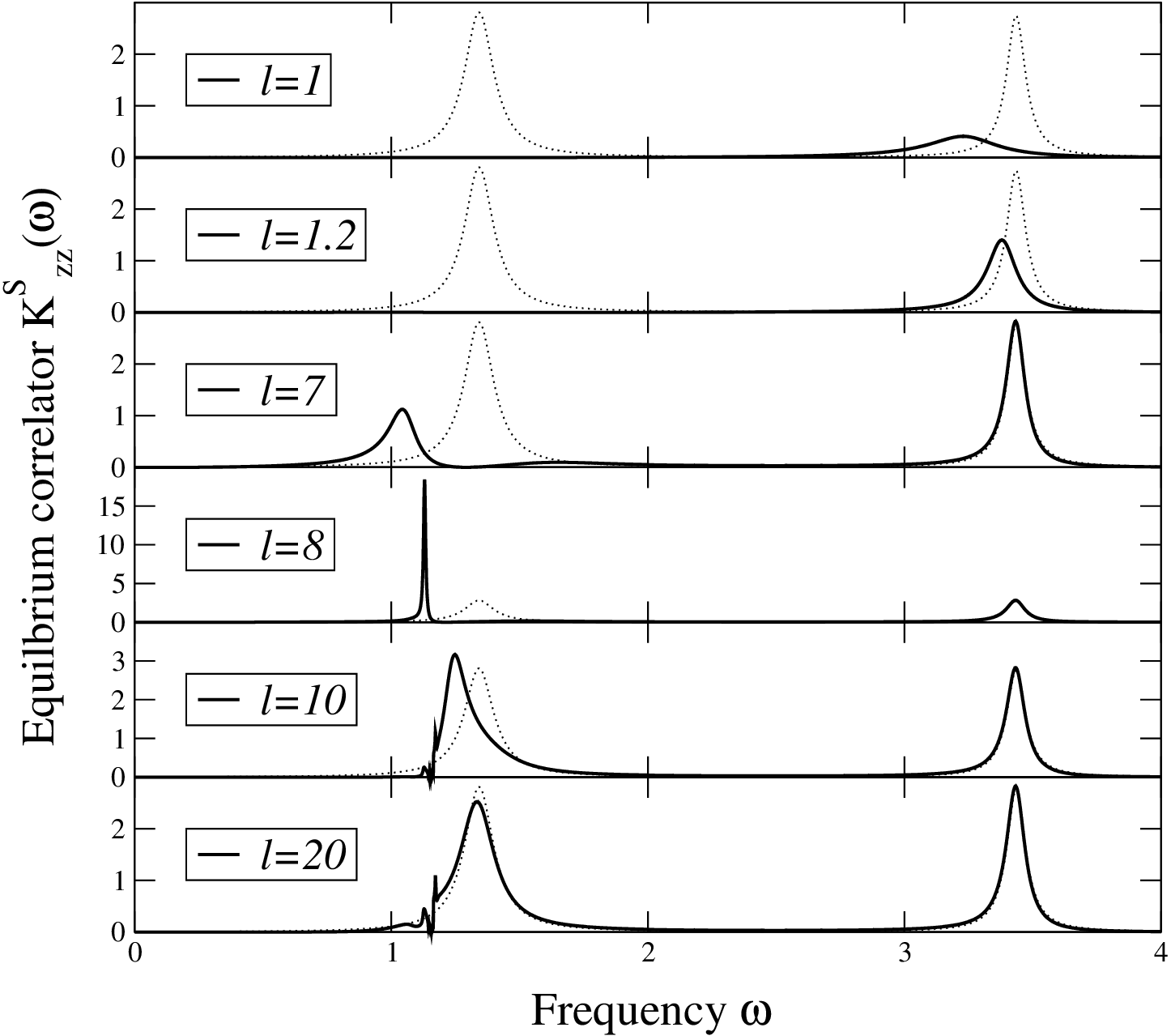}
\caption{\label{fig:ldep}Behavior of the flow using the generator
\(\widetilde{\eta}\): Fourier transform of the equilibrium
correlator for increasing flow parameter \(l\) (solid line), compared
with the final result of a Markoff approximation (dotted line). We see
that the higher modes are decoupled first, i.e., the peak on the
right hand is visible already for \(l=7\). At \(l=20\) the sum
rule is fulfilled with an accuracy of \(\approx 0.5\%\). This result
is better than the Markoff approximation used to calculate the influence of the
linear bath \(F\) on the system \(S+B\), which should lead to
good results for \(\alpha/(2\pi)=0.01\). Naturally, the sum rule is  
not valid at the beginning of the flow. The parameters are: \(\Delta=1\),
\(\Delta_{S}=1.2\), 
\(g=1\), \(\alpha/(2\pi)=0.01\), and \(N=8000\).}
\end{figure}

If we choose the generator \(\widetilde{\eta}\) leading to the
diagonal fixed-point Hamiltonian \(\widetilde{H}_{0}\), which is
scale-independent, then, since no asymptotic scale is present, the
flow equations will first decouple the high-energy modes and then the
low-energy modes. The same feature is found for the correlation
functions, see Fig.~\ref{fig:ldep}. They are first determined for high
energies and the low-energy behavior is calculated last. The frequency
regions of interest are centered on the peaks which appear at the
shifted transition frequencies of the four-level system. The spectral
function around these resonances is determined by a stable flow away
from the asymptotic regime, which must be treated in a different way
\cite{kehrein,stauber,asymp}. In practice, the flow equation can only
be integrated up to \(l^{\ast}\approx(\delta\omega)^{-2}\). We find
that the integration can be stopped for much lower values; the system
is decoupled with high accuracy for \(l>20\), see again
Fig.~\ref{fig:ldep}.

\begin{figure}
\includegraphics[width=80mm]{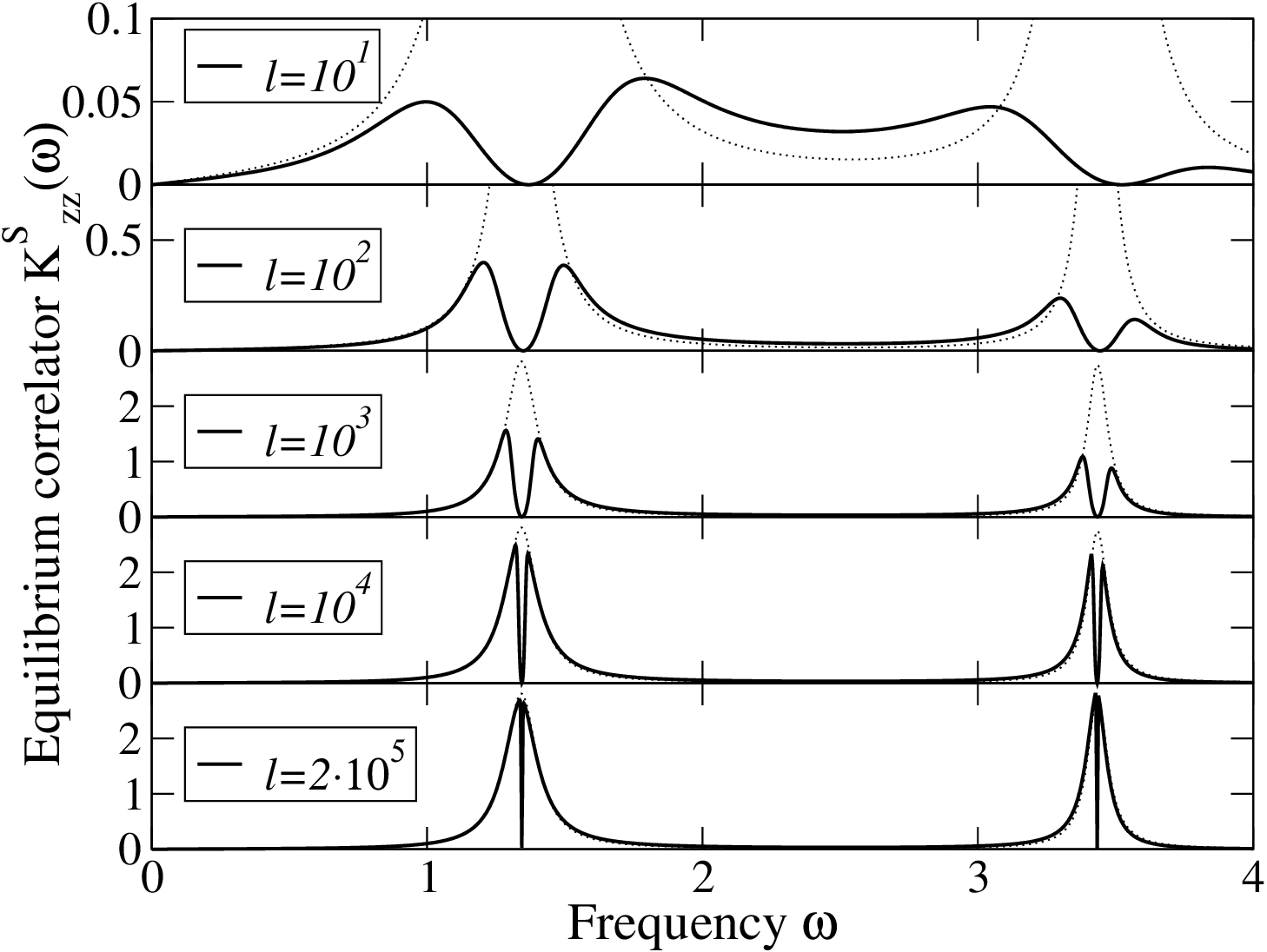}
\caption{\label{fig:nldep}Behavior of the flow using the generator
\(\eta\): Fourier transform of the equilibrium correlator for
increasing flow parameter \(l\) (solid line), compared with the final
result of a Markoff approximation (dotted line). We see that the
frequency regions around the peaks are decoupled at last. The sum rule
is not valid at the beginning of the flow, but finally it is fulfilled
with an accuracy better than \(\approx 0.1\%\). The parameters
are: \(\Delta=1\), \(\Delta_{S}=1.2\), \(g=1\),
\(\alpha/(2\pi)=0.01\), and \(N=2000\).}
\end{figure}

For the other choice of the generator, \(\eta\), we find a different
behavior of the flow. The flow equations first decouple the frequency
regions around the peaks. The peaks contain gaps which are closed only
slowly with increasing \(l\), see Fig.~\ref{fig:nldep}. Therefore, in
contrast to the generator \(\widetilde{\eta}\) the flow equations have
to be integrated up to much higher values \(l\approx 10^5\).

For both choices of the generator there are regions which are
decoupled faster than others. Let us discuss the consequences for the
numerics. The coupling constants of the Hamiltonian,
\(\lambda_{k}^{\alpha\beta}\), and \(\kappa_{k}^{\alpha\beta}\) decay
while the flow parameter \(l\) is varied from zero to infinity. If we
take a closer look at the flow equations, we observe that the
differential equations for a certain mode \(k\) become trivial as soon
as the \(\lambda_{k}^{\alpha\beta}\) and \(\kappa_{k}^{\alpha\beta}\)
are approximately zero, e.g. for the generator \(\widetilde{\eta}\)
the higher modes are first decoupled. In the end, changes take place 
only in a low-frequency region, see
Figs.~\ref{fig:ldep},~\ref{fig:nldep}. This is why, in practice, all
the differential equations outside of this region can be replaced by
trivial ones in the program. This accelerates the calculation
significantly.

The correlation function is given as a bilinear form in the
\(\mu_{k}^{\alpha\beta}\) and \(\nu_{k}^{\alpha\beta}\). The
coefficients \(\mu_{k}^{\alpha\beta}\) and \(\nu_{k}^{\alpha\beta}\)
turn out to be zero at certain points. For the special case of the
nonlinear bath with \(\epsilon_{S}=0\), only \(\mu_{k}^{11}\) remains,
which vanishes (intersects the frequency axis) at different points
\(\omega\). This feature is contained in the flow equations and was
observed in Ref.~\onlinecite{kehrein} for the dissipative two-state
system. The correlation function, bilinear in \(\mu_{k}^{11}\),
therefore has unphysical zeroes, which constitute a finite-size
effect, only disappearing in the thermodynamic limit.

There are different possibilities to address this problem. In
Ref.~\onlinecite{kehrein} it was observed that for a certain value of
the flow parameter, the decoupled two-state system behaved like a
dissipative harmonic oscillator. Furthermore, it was shown that for the
dissipative harmonic oscillator there exists a conserved quantity
which could be added at a certain value of \(l\) in order to complete
the correlation function. Another possible way to deal with this
problem is to integrate to a large value of \(l\), such that the
higher frequencies of the right peak are decoupled, as in
Fig.~\ref{fig:ldep}, and to continue the integration in a restricted
frequency range with a higher resolution of the bath modes. Finally,
only a tiny gap is left which can then be closed by hand.

Using the generator \(\widetilde{\eta}\) spikes appear
around this singularity, which are due to the
finite number of
bath modes \(N\), i.e., we assume that the phenomena will disappear
for increasing \(N\). These spikes, generated via an amplifying
effect, are not completely understood. Probably, they occur always if
the change in the correlation function happens too abruptly. This
suggestion is motivated by the observation that the same feature
appears, if the cutoff is chosen too close to the characteristic
frequency of the system. There, we find that, due to the intrinsic
properties of the differential-equation system, a spike appears at the
cutoff frequency. The system always needs a certain frequency range to
spread out smoothly.

The sum rules are fulfilled with an accuracy of \(\sim 0.5\%\) using
\(\widetilde{\eta}\) and \(\sim 0.1\%\) using the generator \(\eta\),
when not otherwise mentioned.

The bath cutoff was chosen to be \(\omega_{c}=5\), when not otherwise
mentioned. The results depend on the choice of the bath cutoff
\(\omega_{c}\). However, the qualitative behavior is independent of
the cutoff and remains the same for different large-enough
\(\omega_{c}\).

Flow equations give the best results for small coupling
\(\alpha\). This does not mean that the approach is perturbative in
the usual sense, but the neglected terms in the truncation procedure
are smaller in the low-coupling limit. The flow equations for our
specific system obtained with the generator \(\widetilde{\eta}\) were
observed
to give accurate results up to \(\alpha/(2\pi)=0.1\) and
the generator \(\eta\) was used up to \(\alpha/(2\pi)=0.08\).

\section{Comparison of nonlinear with linear bath}
\label{flow:sec:complinnonlin}

The relevant parameters in our model are \( \epsilon _{S},\, \Delta
_{S},\, \Delta, \) the coupling strength \( g \), and the coupling
strength \( \alpha \) to the linear bath $F$. We choose the time scale
such that \( \Delta \equiv 1 \). Furthermore, the results discussed in
the following have been calculated for
\(\Delta_{S}=1.2\), close to \(1\), to keep the decay
strong without being in resonance. 
We set the temperature to
zero (\(T=0\)) and vary the two parameters \(\alpha\) and \(g\).

To begin our discussion, we note some generic features of the results
obtained for the two approaches: linear and nonlinear bath. Since
$K_{zz}^S(\omega)$ is essentially the Fourier transform of the
relaxational dynamics, it consists of several ``Lorentzian like''
peaks. Their number is constrained to be less than the maximum number
of $6$ transition frequencies for $S+B$ in the case of the
nonlinear-bath system. In practice, degeneracies between transition
frequencies and selection rules reduce that number, e.g., to $2$ for
the unbiased case. More peaks, than the one transition frequency expected
for a two level system, are observed in the case of the linear-bath
approximation. It turns out that the second peak is induced by the
peaked bath spectral density. To summarize, we have found for the
unbiased case two peaks for the linear and the nonlinear bath. One 
can associate two different decoherence times with the two peaks. Both
approaches show a similar behavior for increasing coupling \(g\),
where we observe a larger separation between the two peaks. For the
unbiased case \(\epsilon_{S}\neq 0\), we expect at least three peaks
for weak coupling \(\alpha\). Additional possible zero-frequency
delta-distribution contributions show up for the biased case, due to
the definition of the correlation functions. 
Moreover, in contrast to
the high-temperature results, no ``pure'' relaxation is seen in the
biased case.

\begin{figure}
\includegraphics[width=80mm]{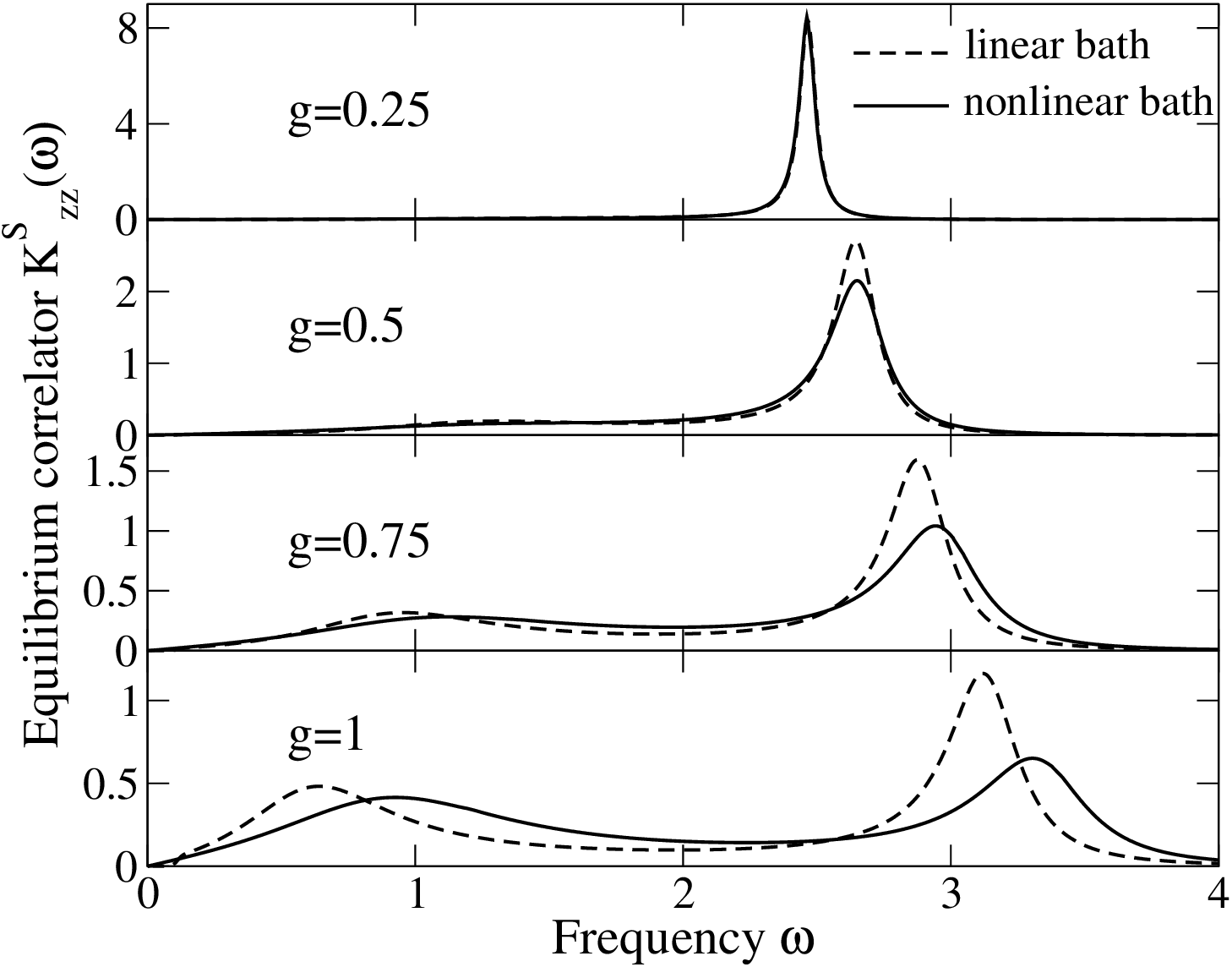}
\caption{\label{fig:alpha0-1}Fourier transform \protect\(
K^{S}_{zz}(\omega )\protect \) of the equilibrium correlator of
\protect\( \hat{\sigma }^{S}_{z}(t) \protect \), for different values
of the system-bath coupling \(g =0.25, 0.5, 0.75\), and \(1\) from
topmost to lowest graph.  The values of the other parameters are:
\protect\( \Delta=1,\, \Delta _{S}=1.2, \alpha/(2\pi)=0.1 \), and
\protect\( \epsilon _{S}=0\protect \). Solid line: nonlinear bath,
\(N=2000\), \(l=8\cdot 10^4\) maximal, and the generator \(\eta\) was
used. Dashed line: linear bath, \(N=2000\), \(l=1000\), and the
generator \(\widetilde{\eta}\) was used.}
\end{figure}

\begin{figure}
\includegraphics[width=80mm]{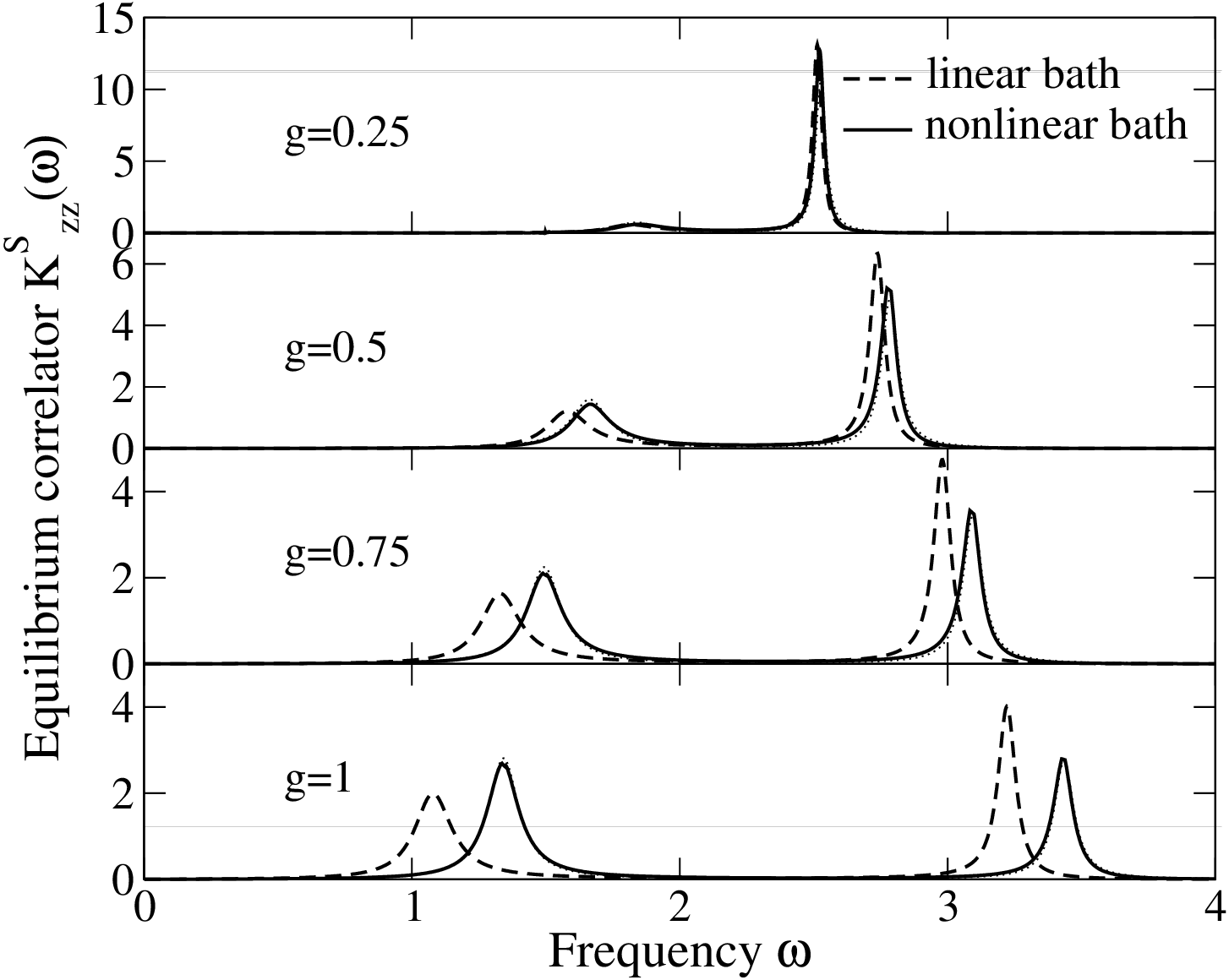}
\caption{\label{fig:alpha0-01}Fourier transform \protect\(
K^{S}_{zz}(\omega )\protect \) of the equilibrium correlator of
\protect\( \hat{\sigma }^{S}_{z}(t)\protect \), for different values
of the system-bath coupling \(g=0.25, 0.5, 0.75\), and \(1\) from
topmost to lowest graph.  The values of the other parameters are:
\protect\( \Delta=1,\, \Delta _{S}=1.2, \alpha/(2\pi)=0.01 \), and
\protect\( \epsilon _{S}=0\protect \). Solid line: nonlinear bath,
\(N=2000\), \(l=2\cdot 10^5\) maximal, and the generator \(\eta\) was
used. Dashed line: linear bath, \(N=2000\), \(l=1000\), and the
generator \(\widetilde{\eta}\) was used. Dotted line (falls together
almost perfectly with the solid line): Markoff approximation used to
calculate the influence of the linear bath \(F\) on the system
\(S+B\).}
\end{figure}

Let us now focus on the unbiased case, i.e., \(\epsilon_{S}=0\). In
the limit of weak coupling, $g\rightarrow 0$, all that remains is a
broadened peak at the transition frequency $2\Delta_S$ of system $S$
alone. In that limit, the results for the two approaches coincide, as
expected. This is visible in the topmost plots of Figs.~\ref{fig:alpha0-1} and
\ref{fig:alpha0-01}, where the Fourier transform of the equilibrium
correlator is shown for increasing system-bath coupling \(g\), while
\(\alpha/(2\pi)=0.05, 0.01\) is kept constant. We expect that for
small \(g\) the nonlinear and the linear-bath results fall together,
i.e., in this limit a weak-coupling approximation \cite{Gardiner00a}
or a Markoff approximation \cite{Blum96a} are valid and one can
therefore not distinguish between linear and nonlinear
bath. Furthermore, the bath spectral density was chosen such that the
two approaches agree for small enough coupling.

With increasing $g$, the peaks are broadened and shifted, and
additional peaks may appear (see Fig.~\ref{fig:alpha0-1}. Indeed, the
most notable difference from a master equation used for coupling of
\(S\) and the nonlinear bath is the appearance of a second peak around
the transition frequency $2\Delta$ of the two-level fluctuator $B$. In
this way, the power spectrum of the bath fluctuations appears in the
short-time behavior of the correlator of the system $S$. This behavior
cannot be captured by a master equation, where only one peak is
present. Here, for the parameters chosen the linear and nonlinear
bath agree nicely for \(g=0.25\), rather to our surprise, very well
up to \(g=0.5\). For \(g>0.5\) the deviations become
significant. Increasing $g$ leads to a frequency shift and a change in
the width of the ``original'' peak at $2\Delta_S$.
These changes are due to the change in eigenfrequencies and eigenvectors
of the combined system $S+B$. 
\begin{figure}
\includegraphics[width=80mm]{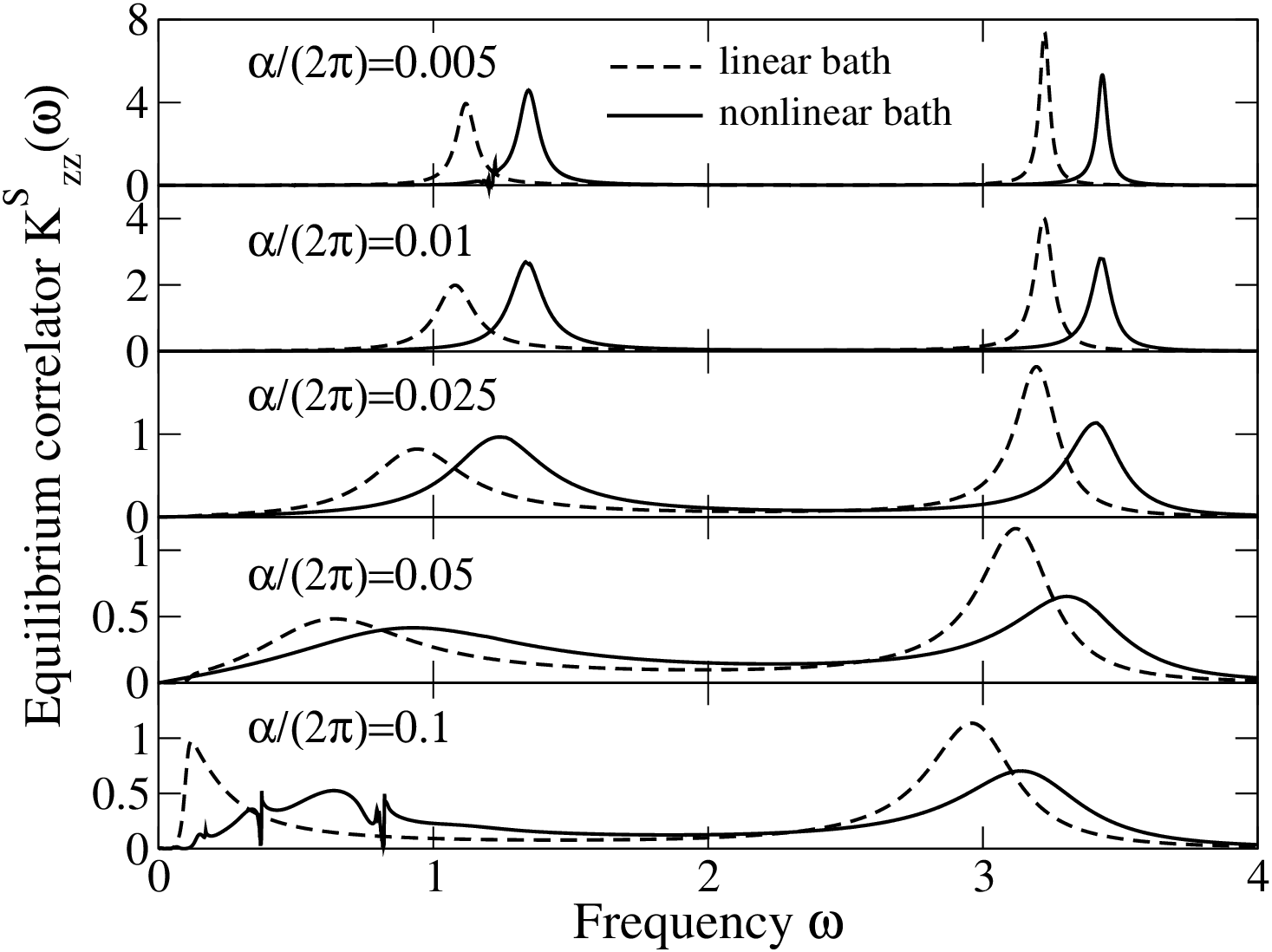}
\caption{\label{fig:acomp}Fourier transform \protect\(
K^{S}_{zz}(\omega )\protect \) of the equilibrium correlator of
\protect\(\hat{\sigma }^{S}_{z}(t)\protect\), for different structured
nonlinear baths with \(\alpha/(2\pi)=0.005, 0.01, 0.025, 0.05\), and
\(0.1\) from topmost to lowest graph. The values of the other
parameters are: \protect\( \Delta=1,\, \Delta _{S}=1.2, g=1\), and
\protect\(\epsilon _{S}=0\protect\). Nonlinear bath: solid line,
linear bath: dashed line. Dotted line: Markoff approximation. In
Sec.~\ref{flow:sec:numdet} we discuss the singularities appearing in
the correlators, e.g., in the lowest graph at frequency
\(\omega\approx 0.8\) and the topmost graph. These two graphs were
obtained using the generator \(\widetilde{\eta}\), \(N=5000\), and
\(l=500\). The rest of the graphs were calculated with the generator
\(\eta\), \(N=2000\), and \(l=2\cdot 10^5\) maximal.}
\end{figure}
If we keep \(g=1\) constant and compare the results for different
increasing \(\alpha\), see Fig.~\ref{fig:acomp}, the differences
between linear and nonlinear bath depend on the ratio \(g/\alpha\). A
small \(\alpha/(2\pi)=0.005\) or \(\alpha/(2\pi)=0.01\) corresponds to
a structured bath (Fig.~\ref{fig:bbcorrb}) and the results show that
the peak shape is similar for nonlinear and linear baths, but the peak
position is different. We stress that for higher coupling, e.g.,
\(\alpha/(2\pi)=0.05\) and \(\alpha/(2\pi)=0.1\), the nonlinear bath acquires a new structure and
the shape of the peak deviates significantly from the linear
bath. Thus, the qualitative differences between linear and nonlinear
baths are smaller for a structured bath.

Since the linear bath takes the full bath spectrum \(\left\langle BB
\right\rangle_{\omega }\) as input, this spectrum may show up in the
result for the system correlator $K_{zz}^S(\omega)$, as is indeed the
case. Figures~\ref{fig:alpha0-1} and \ref{fig:alpha0-01} demonstrate
that this effect is independent of the $\alpha$ values considered here, in
contrast to the infinite temperature limit~\cite{Gassmann02}, where
this effect was most pronounced for small $\gamma$, where the bath
spectrum has a relatively sharp structure. To emphasize this point, we
note that even the asymmetric shape of the bath spectral density
\(\left\langle B B\right\rangle_{\omega}\) is mapped to the
equilibrium correlator, see, e.g., Fig.~\ref{fig:alpha0-1} compared
with Fig.~\ref{fig:bbcorrb}. This is an effect of the non-ohmic
structure of the bath, which could not be observed for an ohmic linear
bath. It can further only be observed when \(S\) and \(B\) are in
resonance, i.e. for low frequencies the bath spectral density behaves
ohmically. The second peak, which appears for higher coupling \(g\),
is therefore determined by the bath correlation function
\(\left\langle B B\right\rangle_{\omega}\) with one peak in the
vicinity of the system transition frequency and another given by the
energy scale \(2\Delta _{S}=2.4\). For increasing system-bath coupling
\(g\), both peaks are shifted away from each other. Within the
linear-bath approximation, the peak around \(2\Delta _{S}\) is shifted
to higher frequencies to a lesser extent and the second peak around
\(2\Delta\) is shifted to lower frequencies, more pronounced compared
to the nonlinear bath.

The results of the linear and nonlinear bath agree well up to
intermediate coupling strengths, as discussed
previously. Nevertheless, there are discrepancies: In particular, the
shifts are different. In contrast to infinite temperature, varying the
parameters \(g\) and \(\alpha\) leads to a visible shift of the peaks
for the linear bath. Furthermore, the peaks become wider and more
asymmetric, but not as pronounced as the peak around \(2\Delta_{S}\)
for infinite temperature. For higher values of $\alpha$, the linear
bath, in general, shows less structure than the actual nonlinear bath,
however, the asymmetric shape of the peaks is mimicked to some
extent.


\begin{figure}
\includegraphics[width=80mm]{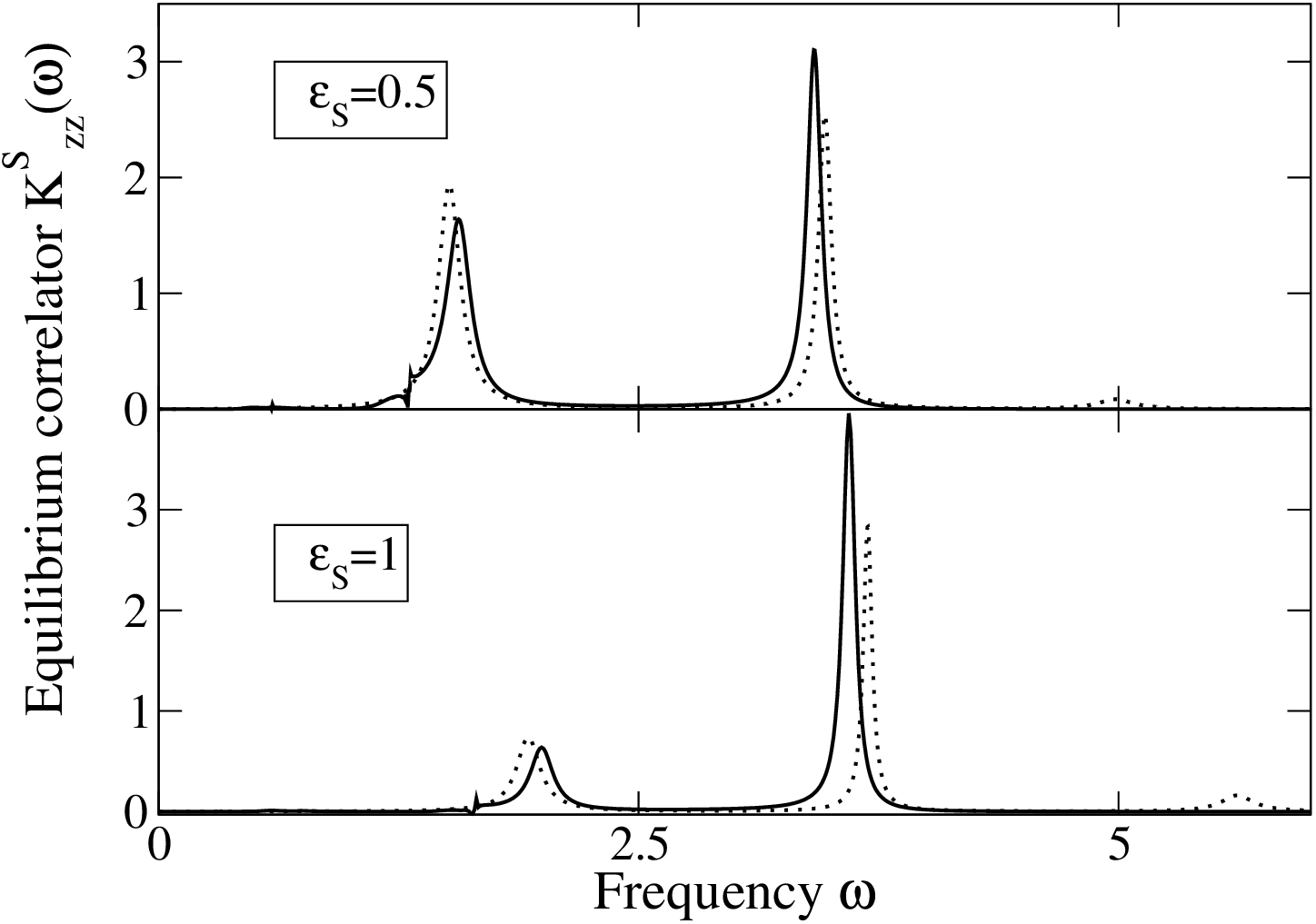}
\caption{\label{fig:epspic}Fourier transform 
\protect\( K^{S}_{zz}(\omega )\protect \)
of the equilibrium correlator of 
\protect\( \hat{\sigma }^{S}_{z}(t)\protect \),
for two different values of the bias \(\epsilon_{S}=0.5\) topmost
\(\epsilon_{S}=1\) lowest graph. The values of the other
parameters are: \protect\( \Delta=1,\, \Delta _{S}=1.2,
\alpha/(2\pi)=0.01 \), \(g=1\),
\(\omega_{C}=10\), \(l=100\), and the generator \(\widetilde{\eta}\) was
used.
For \(\epsilon_{S}=0.5\), \(N=2000\) bath modes were used, the height
of the delta peak at zero frequency is
\(0.229\) and the sum rule leads to a value \(1.048\).
For \(\epsilon_{S}=1\), \(N=1000\) bath modes were used, the height
of the delta peak at zero frequency is
\(0.544\) and the sum rule leads to a value \(1.149\). Solid line:
nonlinear bath, dotted line: Markoff approximation used to calculate
the influence of the linear bath \(F\) on the system \(S+B\).}
\end{figure}

Finally, to close this section let us turn to the biased case
\(\epsilon_{S}\neq 0\). Figure~\ref{fig:epspic} shows the equilibrium
correlator of \(\hat{\sigma }^{S}_{z}(t)\) for two different values of
the bias: \(\epsilon_{S}=0.5\) and \(\epsilon_{S}=1\). For comparison,
the Markoff approximation is shown, which should lead to reasonable
results for \(\alpha/(2\pi)=0.01\), but the agreement between the two
approaches is worse than for the unbiased case. We note that the flow
equations do not describe the third peak. What is the reason for this?
Flow equations are equivalent to a unitary basis change. Therefore, we
expect the same results for each unitarily equivalent Hamiltonian. As
soon as approximations are made, things do change, since, depending on
the representation of the Hamiltonian, different terms have different
significance. Our approximation scheme leads to better results for the
unbiased case. This was also seen in Ref.~\onlinecite{stauber} for the
spin-boson problem, where shifts of the bosonic modes were introduced
and tuned to an optimal point in order to deal with this problem. We
leave such an analysis for the dissipative four-state system for
future work.

\section{Conclusions}
\label{flow::sec:conclusion}
The first main result of this work is the derivation of flow equations
for a general dissipative four-level system. With this framework, not
only the correlation functions of our nonlinear bath model can be
studied, but correlation functions for any four-state system coupled
linearly to an oscillator bath.
An additional key aspect is that there is no restriction for this
bath to have an ohmic bath spectral density.

As an application of these general flow equations, a model of a
nonlinear bath was discussed, consisting of a single two-level system
subject to a linear oscillator bath. Its effect on another two-level
system at zero temperature has been analyzed and compared with the
effect of a linear oscillator bath,
Many numerical results for various special cases have been
obtained and discussed. For the flow equations of the nonlinear bath,
two different generators have been used and the behavior of the
corresponding flows has been analyzed.

For small system-bath coupling the equilibrium correlator contains
only one peak which can be obtained using a Markoff approximation. In
contrast the linear bath can describe the second peak appearing for
larger coupling strengths. At least two frequencies are therefore
present in the time evolution of the two-state system, and
furthermore, the decoherence is strongly increased if the system and
bath are in resonance. As expected, the linear-bath approximation
fails for the regime of large system-bath coupling. In that regime the
linear bath may undoubtedly represent a good approximation to our
actual nonlinear bath up to couplings on the order of half of the
system energy scale. In the strong-coupling regime, on the order of
the system energy scale, the agreement of the peak shapes is
qualitatively better, when the bath spectrum has a strongly peaked
structure. Here, discrepancies between the linear and the original
nonlinear bath became clearly visible. Although we have only discussed
the simplest example of a nonlinear bath, it is likely that the
linear-bath approximation might lead to reasonable results in the
intermediate-coupling regime, also for other, possibly more
sophisticated systems.

\begin{acknowledgments}
We would like to thank S. Kehrein for many crucial hints and
illuminating discussions, and for reading the manuscript. We would
also like to acknowledge S. Gassmann and F. Marquardt. Our work was
supported by the SKORE-A program, the Swiss NSF, and the NCCR
Nanoscience.
\end{acknowledgments}

\appendix
\begin{widetext}
\section{The flow equations for the
Hamiltonian}\label{flow:sec:flow-ham}
Below we list the result for the dissipative four-state
system discussed in Sec.~\ref{flow:sec:nonlinear}. A sum over
\(i,j,l,p,s,r\), and \(t\) from one to three is suppressed in the formulas.
\begin{eqnarray}\nonumber
\frac{d\Delta^{0
0}}{dl}&=&4\sum_{k}\Big\{-\frac{1}{2}\Big(\lambda_{k}^{0
0}\lambda_{k}^{0 0}+\lambda_{k}^{i j}\lambda_{k}^{i j}+\lambda_{k}^{0
j}\lambda_{k}^{0 j}+\lambda_{k}^{i 0}\lambda_{k}^{i 0} +\kappa_{k}^{0
0}\kappa_{k}^{0 0}+\kappa_{k}^{i j}\kappa_{k}^{i j}+\kappa_{k}^{0
j}\kappa_{k}^{0 j}+\kappa_{k}^{i 0}\kappa_{k}^{i 0}\Big)\omega_{k}
\\\nonumber
&-&(\Delta^{0p}\kappa_{k}^{0l}+\Delta^{sp}\kappa_{k}^{sl})\epsilon_{plj}\lambda_{k}^{0j}
-(\Delta^{s0}\kappa_{k}^{l0}+\Delta^{sp}\kappa_{k}^{lp})\epsilon_{sli}\lambda_{k}^{i0}
-(\Delta^{0p}\kappa_{k}^{il}\epsilon_{plj}+\Delta^{s0}\kappa_{k}^{lj}\epsilon_{sli}
+\Delta^{ip}\kappa_{k}^{0l}\epsilon_{plj}+\Delta^{sj}\kappa_{k}^{l0}\epsilon_{sli})\lambda_{k}^{ij}
\\\nonumber &+&(\Delta^{sp}\lambda_{k}^{sl}
+\Delta^{0p}\lambda_{k}^{0l})\epsilon_{plj}\kappa_{k}^{0j}
+(\Delta^{s0}\lambda_{k}^{l0}
+\Delta^{sp}\lambda_{k}^{lp})\epsilon_{sli}\kappa_{k}^{i0}
+(\Delta^{0p}\lambda_{k}^{il}\epsilon_{plj}+\Delta^{s0}\lambda_{k}^{lj}\epsilon_{sli}+\Delta^{ip}\lambda_{k}^{0l}\epsilon_{plj}+\Delta^{sj}\lambda_{k}^{l0}\epsilon_{sli})\kappa_{k}^{ij}\Big\}
\\\nonumber \frac{d\Delta^{0
n}}{dl}&=&4\sum_{k}\Big\{\big[-\lambda_{k}^{i 0}\lambda_{k}^{i
n}-\lambda_{k}^{0 0}\lambda_{k}^{0 n}-\kappa_{k}^{i 0}\kappa_{k}^{i
n}-\kappa_{k}^{0 0}\kappa_{k}^{0 n} +\frac{1}{2}(\kappa_{k}^{0
i}\lambda_{k}^{0 l}+\kappa_{k}^{j i}\lambda_{k}^{j l}-\lambda_{k}^{0
i}\kappa_{k}^{0 l}-\lambda_{k}^{j i}\kappa_{k}^{j l})\epsilon_{i l
n}(2 n_{k}+1)\big]\omega_{k} \\\nonumber
&-&\big[(\Delta^{sp}\lambda_{k}^{sl}
+\Delta^{0p}\lambda_{k}^{0l})\lambda_{k}^{0r}\epsilon_{plj}\epsilon_{jrn}
+(\Delta^{0p}\lambda_{k}^{il}\epsilon_{plj}+\Delta^{s0}\lambda_{k}^{lj}\epsilon_{sli}
+\Delta^{ip}\lambda_{k}^{0l}\epsilon_{plj}+\Delta^{sj}\lambda_{k}^{l0}\epsilon_{sli})\lambda_{k}^{ir}\epsilon_{jrn}
\\\nonumber &+&(\Delta^{0p}\kappa_{k}^{0l}
+\Delta^{sp}\kappa_{k}^{sl})\kappa_{k}^{0r}\epsilon_{plj}\epsilon_{jrn}
+(\Delta^{0p}\kappa_{k}^{il}\epsilon_{plj}+\Delta^{s0}\kappa_{k}^{lj}\epsilon_{sli}
+\Delta^{ip}\kappa_{k}^{0l}\epsilon_{plj}+\Delta^{sj}\kappa_{k}^{l0}\epsilon_{sli})\kappa_{k}^{ir}\epsilon_{jrn}\big](2n_{k}+1)
\\\nonumber &-&(\Delta^{0p}\kappa_{k}^{0l}
+\Delta^{sp}\kappa_{k}^{sl})\lambda_{k}^{00}\epsilon_{pln}
-(\Delta^{s0}\kappa_{k}^{l0}
+\Delta^{sp}\kappa_{k}^{lp})\lambda_{k}^{in}\epsilon_{sli}
-(\Delta^{0p}\kappa_{k}^{il}\epsilon_{pln}+\Delta^{s0}\kappa_{k}^{ln}\epsilon_{sli}
+\Delta^{ip}\kappa_{k}^{0l}\epsilon_{pln}+\Delta^{sn}\kappa_{k}^{l0}\epsilon_{sli})\lambda_{k}^{i0}
\\\nonumber
&+&(\Delta^{sp}\lambda_{k}^{sl}+\Delta^{0p}\lambda_{k}^{0l})\kappa_{k}^{00}\epsilon_{pln}
+(\Delta^{s0}\lambda_{k}^{l0}+\Delta^{sp}\lambda_{k}^{lp})\kappa_{k}^{in}\epsilon_{sli}
+(\Delta^{0p}\lambda_{k}^{il}\epsilon_{pln}+\Delta^{s0}\lambda_{k}^{ln}\epsilon_{sli}+\Delta^{ip}\lambda_{k}^{0l}\epsilon_{pln}+\Delta^{sn}\lambda_{k}^{l0}\epsilon_{sli})\kappa_{k}^{i0}\Big\}
\\\nonumber \frac{d\Delta^{n
0}}{dl}&=&4\sum_{k}\Big\{\big[-\lambda_{k}^{0 i}\lambda_{k}^{n
i}-\lambda_{k}^{0 0}\lambda_{k}^{n 0}-\kappa_{k}^{0 i}\kappa_{k}^{n
i}-\kappa_{k}^{0 0}\kappa_{k}^{n 0} +\frac{1}{2}(\kappa_{k}^{i
0}\lambda_{k}^{l 0}+\kappa_{k}^{i j}\lambda_{k}^{l j}-\lambda_{k}^{i
0}\kappa_{k}^{l 0}-\lambda_{k}^{i j}\kappa_{k}^{l j})\epsilon_{i l
n}(2 n_{k}+1)\big]\omega_{k} \\\nonumber
&-&\big[(\Delta^{s0}\lambda_{k}^{l0}
+\Delta^{sp}\lambda_{k}^{lp})\lambda_{k}^{r0}\epsilon_{sli}\epsilon_{irn}
+(\Delta^{0p}\lambda_{k}^{il}\epsilon_{plj}+\Delta^{s0}\lambda_{k}^{lj}\epsilon_{sli}+\Delta^{ip}\lambda_{k}^{0l}\epsilon_{plj}+\Delta^{sj}\lambda_{k}^{l0}\epsilon_{sli})\lambda_{k}^{rj}\epsilon_{irn}
\\\nonumber
&+&(\Delta^{s0}\kappa_{k}^{l0}+\Delta^{sp}\kappa_{k}^{lp})\kappa_{k}^{r0}\epsilon_{sli}\epsilon_{irn}
+(\Delta^{0p}\kappa_{k}^{il}\epsilon_{plj}+\Delta^{s0}\kappa_{k}^{lj}\epsilon_{sli}
+\Delta^{ip}\kappa_{k}^{0l}\epsilon_{plj}+\Delta^{sj}\kappa_{k}^{l0}\epsilon_{sli})\kappa_{k}^{rj}\epsilon_{irn}\big](2n_{k}+1)
\\\nonumber
&-&(\Delta^{0p}\kappa_{k}^{nl}\epsilon_{plj}+\Delta^{s0}\kappa_{k}^{lj}\epsilon_{sln}
+\Delta^{np}\kappa_{k}^{0l}\epsilon_{plj}+\Delta^{sj}\kappa_{k}^{l0}\epsilon_{sln})\lambda_{k}^{0j}
-(\Delta^{s0}\kappa_{k}^{l0}+\Delta^{sp}\kappa_{k}^{lp})\lambda_{k}^{00}\epsilon_{sln}
-(\Delta^{0p}\kappa_{k}^{0l}+\Delta^{sp}\kappa_{k}^{sl})\lambda_{k}^{nj}\epsilon_{plj}
\\\nonumber
&+&(\Delta^{0p}\lambda_{k}^{nl}\epsilon_{plj}+\Delta^{s0}\lambda_{k}^{lj}\epsilon_{sln}+\Delta^{np}\lambda_{k}^{0l}\epsilon_{plj}+\Delta^{sj}\lambda_{k}^{l0}\epsilon_{sln})\kappa_{k}^{0j}
+(\Delta^{sp}\lambda_{k}^{sl}+\Delta^{0p}\lambda_{k}^{0l})\kappa_{k}^{nj}\epsilon_{plj}
+(\Delta^{s0}\lambda_{k}^{l0}+\Delta^{sp}\lambda_{k}^{lp})\kappa_{k}^{00}\epsilon_{sln}\Big\}
\end{eqnarray}
\begin{eqnarray}\nonumber
\frac{d\Delta^{n m}}{dl}&=&4\sum_{k}\Big\{\Big(-\lambda_{k}^{0 0}\lambda_{k}^{n
m}-\lambda_{k}^{0 m}\lambda_{k}^{n 0}-\kappa_{k}^{0 0}\kappa_{k}^{n
m}-\kappa_{k}^{0 m}\kappa_{k}^{n 0}
+\frac{1}{2}(\lambda_{k}^{i j}\lambda_{k}^{p
l}+\kappa_{k}^{i j}\kappa_{k}^{p
l})\epsilon_{i p n}\epsilon_{j l m}
\\\nonumber
&+&\frac{1}{2}\big[(\kappa_{k}^{i 0}\lambda_{k}^{l m}+\kappa_{k}^{i
m}\lambda_{k}^{l 0}-\lambda_{k}^{i 0}\kappa_{k}^{l m}-\lambda_{k}^{i
m}\kappa_{k}^{l 0})\epsilon_{i l n}
+(\kappa_{k}^{0 i}\lambda_{k}^{n
l}+\kappa_{k}^{n i}\lambda_{k}^{0 l}-\lambda_{k}^{0 i}\kappa_{k}^{n
l}-\lambda_{k}^{n i}\kappa_{k}^{0 l})\epsilon_{i l m}\big](2
n_{k}+1)\Big)\omega_{k}
\\\nonumber
&-&\big[(\Delta^{sp}\lambda_{k}^{sl}+\Delta^{0p}\lambda_{k}^{0l})\lambda_{k}^{nr}\epsilon_{plj}\epsilon_{jrm}
+(\Delta^{0p}\lambda_{k}^{nl}\epsilon_{plj}+\Delta^{s0}\lambda_{k}^{lj}\epsilon_{sln}+\Delta^{np}\lambda_{k}^{0l}\epsilon_{plj}+\Delta^{sj}\lambda_{k}^{l0}\epsilon_{sln})\lambda_{k}^{0r}\epsilon_{jrm}
\\\nonumber
&+&(\Delta^{s0}\lambda_{k}^{l0}+\Delta^{sp}\lambda_{k}^{lp})\lambda_{k}^{rm}\epsilon_{sli}\epsilon_{irn}
+(\Delta^{0p}\lambda_{k}^{il}\epsilon_{plm}+\Delta^{s0}\lambda_{k}^{lm}\epsilon_{sli}+\Delta^{ip}\lambda_{k}^{0l}\epsilon_{plm}+\Delta^{sm}\lambda_{k}^{l0}\epsilon_{sli})\lambda_{k}^{r0}\epsilon_{irn}
\\\nonumber
&+&(\Delta^{0p}\kappa_{k}^{0l}+\Delta^{sp}\kappa_{k}^{sl})\kappa_{k}^{nr}\epsilon_{plj}\epsilon_{jrm}
+(\Delta^{0p}\kappa_{k}^{nl}\epsilon_{plj}+\Delta^{s0}\kappa_{k}^{lj}\epsilon_{sln}
+\Delta^{np}\kappa_{k}^{0l}\epsilon_{plj}+\Delta^{sj}\kappa_{k}^{l0}\epsilon_{sln})\kappa_{k}^{0r}\epsilon_{jrm}
\\\nonumber
&+&(\Delta^{s0}\kappa_{k}^{l0}+\Delta^{sp}\kappa_{k}^{lp})\kappa_{k}^{rm}\epsilon_{sli}\epsilon_{irn}
+(\Delta^{0p}\kappa_{k}^{il}\epsilon_{plm}+\Delta^{s0}\kappa_{k}^{lm}\epsilon_{sli}
+\Delta^{ip}\kappa_{k}^{0l}\epsilon_{plm}+\Delta^{sm}\kappa_{k}^{l0}\epsilon_{sli})\kappa_{k}^{r0}\epsilon_{irn}\big](2n_{k}+1)
\\\nonumber
&-&(\Delta^{0p}\kappa_{k}^{0l}
+\Delta^{sp}\kappa_{k}^{sl})\lambda_{k}^{n0}\epsilon_{plm}
+(\Delta^{0p}\kappa_{k}^{il}\epsilon_{plj}+\Delta^{s0}\kappa_{k}^{lj}\epsilon_{sli}
+\Delta^{ip}\kappa_{k}^{0l}\epsilon_{plj}+\Delta^{sj}\kappa_{k}^{l0}\epsilon_{sli})\lambda_{k}^{tr}\epsilon_{itn}\epsilon_{jrm}
\\\nonumber
&-&(\Delta^{s0}\kappa_{k}^{l0}+\Delta^{sp}\kappa_{k}^{lp})\lambda_{k}^{0m}\epsilon_{sln}
-(\Delta^{0p}\kappa_{k}^{nl}\epsilon_{plm}+\Delta^{s0}\kappa_{k}^{lm}\epsilon_{sln}
+\Delta^{np}\kappa_{k}^{0l}\epsilon_{plm}+\Delta^{sm}\kappa_{k}^{l0}\epsilon_{sln})\lambda_{k}^{00}
\\\nonumber
&+&(\Delta^{sp}\lambda_{k}^{sl}+\Delta^{0p}\lambda_{k}^{0l})\kappa_{k}^{n0}\epsilon_{plm}
-(\Delta^{0p}\lambda_{k}^{il}\epsilon_{plj}+\Delta^{s0}\lambda_{k}^{lj}\epsilon_{sli}+\Delta^{ip}\lambda_{k}^{0l}\epsilon_{plj}+\Delta^{sj}\lambda_{k}^{l0}\epsilon_{sli})\kappa_{k}^{tr}\epsilon_{itn}\epsilon_{jrm}
\\\nonumber
&+&(\Delta^{s0}\lambda_{k}^{l0}+\Delta^{sp}\lambda_{k}^{lp})\kappa_{k}^{0m}\epsilon_{sln}
+(\Delta^{0p}\lambda_{k}^{nl}\epsilon_{plm}+\Delta^{s0}\lambda_{k}^{lm}\epsilon_{sln}+\Delta^{np}\lambda_{k}^{0l}\epsilon_{plm}+\Delta^{sm}\lambda_{k}^{l0}\epsilon_{sln})\kappa_{k}^{00}\Big\}
\end{eqnarray}
\begin{eqnarray}\\\nonumber
\frac{d\lambda_{k}^{0 0}}{dl}&=&-\omega_{k}^2\lambda_{k}^{0
0}+2\sum_{q}\eta_{k q}\lambda_{q}^{0 0}
\\\nonumber
\frac{d\lambda_{k}^{0 n}}{dl}&=&-\omega_{k}^2\lambda_{k}^{0
n}+4(\kappa_{k}^{0 i}\Delta^{0 l}+\kappa_{k}^{j i}\Delta^{j
l})\epsilon_{i l n}\omega_{k}+2\sum_{q}\eta_{k q}\lambda_{q}^{0 n}
\\\nonumber
&-&4(\Delta^{sp}\lambda_{k}^{sl}+\Delta^{0p}\lambda_{k}^{0l})\Delta^{0r}\epsilon_{plj}\epsilon_{jrn}
-4(\Delta^{0p}\lambda_{k}^{il}\epsilon_{plj}+\Delta^{s0}\lambda_{k}^{lj}\epsilon_{sli}+\Delta^{ip}\lambda_{k}^{0l}\epsilon_{plj}+\Delta^{sj}\lambda_{k}^{l0}\epsilon_{sli})\Delta^{ir}\epsilon_{jrn}
\\\nonumber
\frac{d\lambda_{k}^{n 0}}{dl}&=&-\omega_{k}^2\lambda_{k}^{n
0}+4(\kappa_{k}^{i 0}\Delta^{l
0}+\kappa_{k}^{i j}\Delta^{l j})\epsilon_{i l
n}\omega_{k}+2\sum_{q}\eta_{k q}\lambda_{q}^{n 0}
\\\nonumber
&-&4(\Delta^{s0}\lambda_{k}^{l0}
+\Delta^{sp}\lambda_{k}^{lp})\Delta^{r0}\epsilon_{sli}\epsilon_{irn}
-4(\Delta^{0p}\lambda_{k}^{il}\epsilon_{plj}+\Delta^{s0}\lambda_{k}^{lj}\epsilon_{sli}+\Delta^{ip}\lambda_{k}^{0l}\epsilon_{plj}+\Delta^{sj}\lambda_{k}^{l0}\epsilon_{sli})\Delta^{rj}\epsilon_{irn}
\\\nonumber
\frac{d\lambda_{k}^{n m}}{dl}&=&-\omega_{k}^2\lambda_{k}^{n
m}+4\Big((\kappa_{k}^{0 i}\Delta^{n l}+\kappa_{k}^{n i}\Delta^{0 l})\epsilon_{i l m}+(\kappa_{k}^{i 0}\Delta^{l m}+\kappa_{k}^{i m}\Delta^{l 0})\epsilon_{i l n}\Big)\omega_{k}
+2\sum_{q}\eta_{k q}\lambda_{q}^{n m}
\\\nonumber
&-&4(\Delta^{sp}\lambda_{k}^{sl}+\Delta^{0p}\lambda_{k}^{0l})\Delta^{nr}\epsilon_{plj}\epsilon_{jrm}
-4(\Delta^{0p}\lambda_{k}^{nl}\epsilon_{plj}+\Delta^{s0}\lambda_{k}^{lj}\epsilon_{sln}+\Delta^{np}\lambda_{k}^{0l}\epsilon_{plj}+\Delta^{sj}\lambda_{k}^{l0}\epsilon_{sln})\Delta^{0r}\epsilon_{jrm}
\\\nonumber
&-&4(\Delta^{s0}\lambda_{k}^{l0}+\Delta^{sp}\lambda_{k}^{lp})\Delta^{rm}\epsilon_{sli}\epsilon_{irn}
-4(\Delta^{0p}\lambda_{k}^{il}\epsilon_{plm}+\Delta^{s0}\lambda_{k}^{lm}\epsilon_{sli}+\Delta^{ip}\lambda_{k}^{0l}\epsilon_{plm}+\Delta^{sm}\lambda_{k}^{l0}\epsilon_{sli})\Delta^{r0}\epsilon_{irn}
\end{eqnarray}
\begin{eqnarray}\nonumber
\frac{d\kappa_{k}^{0 0}}{dl}&=&-\omega_{k}^2\kappa_{k}^{0 0}-2\sum_{q}\eta_{q k}\kappa_{q}^{0 0}
\\\nonumber
\frac{d\kappa_{k}^{0 n}}{dl}&=&-\omega_{k}^2\kappa_{k}^{0
n}-4(\lambda_{k}^{0 i}\Delta^{0 l}+\lambda_{k}^{j i}\Delta^{j
l})\epsilon_{i l n}\omega_{k}-2\sum_{q}\eta_{q k}\kappa_{q}^{0 n}
\\\nonumber
&-&4(\Delta^{0p}\kappa_{k}^{0l}+\Delta^{sp}\kappa_{k}^{sl})\Delta^{0r}\epsilon_{plj}\epsilon_{jrn}
-4(\Delta^{0p}\kappa_{k}^{il}\epsilon_{plj}+\Delta^{s0}\kappa_{k}^{lj}\epsilon_{sli}
+\Delta^{ip}\kappa_{k}^{0l}\epsilon_{plj}+\Delta^{sj}\kappa_{k}^{l0}\epsilon_{sli})\Delta^{ir}\epsilon_{jrn}
\\\nonumber
\frac{d\kappa_{k}^{n 0}}{dl}&=&-\omega_{k}^2\kappa_{k}^{n
0}-4(\lambda_{k}^{i 0}\Delta^{l
0}+\lambda_{k}^{i j}\Delta^{l j})\epsilon_{i l
n}\omega_{k}-2\sum_{q}\eta_{q k}\kappa_{q}^{n 0}
\\\nonumber
&-&4(\Delta^{s0}\kappa_{k}^{l0}+\Delta^{sp}\kappa_{k}^{lp})\Delta^{r0}\epsilon_{sli}\epsilon_{irn}
-4(\Delta^{0p}\kappa_{k}^{il}\epsilon_{plj}+\Delta^{s0}\kappa_{k}^{lj}\epsilon_{sli}
+\Delta^{ip}\kappa_{k}^{0l}\epsilon_{plj}+\Delta^{sj}\kappa_{k}^{l0}\epsilon_{sli})\Delta^{rj}\epsilon_{irn}
\\\nonumber
\frac{d\kappa_{k}^{n m}}{dl}&=&-\omega_{k}^2\kappa_{k}^{n
m}-4\Big((\lambda_{k}^{0
i}\Delta^{n l}+\lambda_{k}^{n i}\Delta^{0 l})\epsilon_{i l
m}+(\lambda_{k}^{i 0}\Delta^{l m}+\lambda_{k}^{i m}\Delta^{l
0})\epsilon_{i l n}\Big)\omega_{k}-2\sum_{q}\eta_{q k}\kappa_{q}^{n m}
\\\nonumber
&-&4(\Delta^{0p}\kappa_{k}^{0l}+\Delta^{sp}\kappa_{k}^{sl})\Delta^{nr}\epsilon_{plj}\epsilon_{jrm}
-4(\Delta^{0p}\kappa_{k}^{nl}\epsilon_{plj}+\Delta^{s0}\kappa_{k}^{lj}\epsilon_{sln}
+\Delta^{np}\kappa_{k}^{0l}\epsilon_{plj}+\Delta^{sj}\kappa_{k}^{l0}\epsilon_{sli})\Delta^{0r}\epsilon_{jrm}
\\\nonumber
&-&4(\Delta^{s0}\kappa_{k}^{l0}+\Delta^{sp}\kappa_{k}^{lp})\Delta^{rm}\epsilon_{sli}\epsilon_{irn}
-4(\Delta^{0p}\kappa_{k}^{il}\epsilon_{plm}+\Delta^{s0}\kappa_{k}^{lm}\epsilon_{sli}
+\Delta^{ip}\kappa_{k}^{0l}\epsilon_{plm}+\Delta^{sm}\kappa_{k}^{l0}\epsilon_{sli})\Delta^{r0}\epsilon_{irn}
\end{eqnarray}

\section{The \(\eta_{k q}\) coefficients}\label{eta}
\begin{eqnarray}\nonumber
\eta_{k q}&=&\frac{1}{\omega_{q}^2-\omega_{k}^2}\Bigg\{2\omega_{k}\Big[\big(\lambda_{k}^{0 i}\kappa_{q}^{n
l}\epsilon_{i l m}\left\langle\Sigma_{n m}\right\rangle+\lambda_{k}^{i 0}\kappa_{q}^{l
m}\epsilon_{i l n}\left\langle\Sigma_{n m}\right\rangle
+\lambda_{k}^{n j}\kappa_{q}^{0
l}\epsilon_{j l m}\left\langle\Sigma_{n m}\right\rangle
+\lambda_{k}^{i m}\kappa_{q}^{l
0}\epsilon_{i l n}\left\langle\Sigma_{n m}\right\rangle
\\\nonumber
&+&\lambda_{k}^{0 i}\kappa_{q}^{0
l}\epsilon_{i l n}\left\langle\Sigma_{0 n}\right\rangle
+\lambda_{k}^{i 0}\kappa_{q}^{l
0}\epsilon_{i l n}\left\langle\Sigma_{n 0}\right\rangle
+\lambda_{k}^{j i}\kappa_{q}^{j
l}\epsilon_{i l n}\left\langle\Sigma_{0 n}\right\rangle
+\lambda_{k}^{i j}\kappa_{q}^{l
j}\epsilon_{i l n}\left\langle\Sigma_{n 0}\right\rangle\big)\omega_{k}
\\\nonumber
&+&\big(
\lambda_{q}^{0 i}\kappa_{k}^{n
l}\epsilon_{i l m}\left\langle\Sigma_{n m}\right\rangle+\lambda_{q}^{i 0}\kappa_{k}^{l
m}\epsilon_{i l n}\left\langle\Sigma_{n m}\right\rangle
+\lambda_{q}^{n j}\kappa_{k}^{0
l}\epsilon_{j l m}\left\langle\Sigma_{n m}\right\rangle
+\lambda_{q}^{i m}\kappa_{k}^{l
0}\epsilon_{i l n}\left\langle\Sigma_{n m}\right\rangle
\\\nonumber
&+&\lambda_{q}^{0 i}\kappa_{k}^{0
l}\epsilon_{i l n}\left\langle\Sigma_{0 n}\right\rangle
+\lambda_{q}^{i 0}\kappa_{k}^{l
0}\epsilon_{i l n}\left\langle\Sigma_{n 0}\right\rangle
+\lambda_{q}^{j i}\kappa_{k}^{j
l}\epsilon_{i l n}\left\langle\Sigma_{0 n}\right\rangle
+\lambda_{q}^{i j}\kappa_{k}^{l
j}\epsilon_{i l n}\left\langle\Sigma_{n
0}\right\rangle\big)\omega_{q}\Big]
\\
&-&2\omega_{q}\Big[\big(\kappa_{k}^{0 i}\lambda_{q}^{n
l}\epsilon_{i l m}\left\langle\Sigma_{n m}\right\rangle+\kappa_{k}^{i 0}\lambda_{q}^{l
m}\epsilon_{i l n}\left\langle\Sigma_{n m}\right\rangle
+\kappa_{k}^{n j}\lambda_{q}^{0
l}\epsilon_{j l m}\left\langle\Sigma_{n m}\right\rangle
+\kappa_{k}^{i m}\lambda_{q}^{l
0}\epsilon_{i l n}\left\langle\Sigma_{n m}\right\rangle
\\\nonumber
&+&\kappa_{k}^{0 i}\lambda_{q}^{0
l}\epsilon_{i l n}\left\langle\Sigma_{0 n}\right\rangle
+\kappa_{k}^{i 0}\lambda_{q}^{l
0}\epsilon_{i l n}\left\langle\Sigma_{n 0}\right\rangle
+\kappa_{k}^{j i}\lambda_{q}^{j
l}\epsilon_{i l n}\left\langle\Sigma_{0 n}\right\rangle
+\kappa_{k}^{i j}\lambda_{q}^{l
j}\epsilon_{i l n}\left\langle\Sigma_{n 0}\right\rangle\big)\omega_{k}
\\\nonumber
&+&\big(
\kappa_{q}^{0 i}\lambda_{k}^{n
l}\epsilon_{i l m}\left\langle\Sigma_{n m}\right\rangle+\kappa_{q}^{i 0}\lambda_{k}^{l
m}\epsilon_{i l n}\left\langle\Sigma_{n m}\right\rangle
+\kappa_{q}^{n j}\lambda_{k}^{0
l}\epsilon_{j l m}\left\langle\Sigma_{n m}\right\rangle
+\kappa_{q}^{i m}\lambda_{k}^{l
0}\epsilon_{i l n}\left\langle\Sigma_{n m}\right\rangle
\\\nonumber
&+&\kappa_{q}^{0 i}\lambda_{k}^{0
l}\epsilon_{i l n}\left\langle\Sigma_{0 n}\right\rangle
+\kappa_{q}^{i 0}\lambda_{k}^{l
0}\epsilon_{i l n}\left\langle\Sigma_{n 0}\right\rangle
+\kappa_{q}^{j i}\lambda_{k}^{j
l}\epsilon_{i l n}\left\langle\Sigma_{0 n}\right\rangle
+\kappa_{q}^{i j}\lambda_{k}^{l
j}\epsilon_{i l n}\left\langle\Sigma_{n 0}\right\rangle\big)\omega_{q}\Big]
\end{eqnarray}

\begin{eqnarray}\nonumber
&+&4\omega_{k}\Big[
(\Delta^{0p}\kappa_{k}^{0l}+\Delta^{sp}\kappa_{k}^{sl})\kappa_{q}^{nr}\epsilon_{plj}\epsilon_{jrm}\left\langle\Sigma_{nm}\right\rangle
+(\Delta^{0p}\kappa_{k}^{nl}\epsilon_{plj}+\Delta^{s0}\kappa_{k}^{lj}\epsilon_{sln}
+\Delta^{np}\kappa_{k}^{0l}\epsilon_{plj}+\Delta^{sj}\kappa_{k}^{l0}\epsilon_{sln})\kappa_{q}^{0r}\epsilon_{jrm}\left\langle\Sigma_{nm}\right\rangle
\\\nonumber
&+&(\Delta^{s0}\kappa_{k}^{l0}+\Delta^{sp}\kappa_{k}^{lp})\kappa_{q}^{rm}\epsilon_{sli}\epsilon_{irn}\left\langle\Sigma_{nm}\right\rangle
+(\Delta^{0p}\kappa_{k}^{il}\epsilon_{plm}+\Delta^{s0}\kappa_{k}^{lm}\epsilon_{sli}
+\Delta^{ip}\kappa_{k}^{0l}\epsilon_{plm}+\Delta^{sm}\kappa_{k}^{l0}\epsilon_{sli})\kappa_{q}^{r0}\epsilon_{irn}\left\langle\Sigma_{nm}\right\rangle
\\\nonumber
&+&(\Delta^{0p}\kappa_{k}^{0l}+\Delta^{sp}\kappa_{k}^{sl})\kappa_{q}^{0r}\epsilon_{plj}\epsilon_{jrn}\left\langle\Sigma_{0n}\right\rangle
+(\Delta^{0p}\kappa_{k}^{il}\epsilon_{plj}+\Delta^{s0}\kappa_{k}^{lj}\epsilon_{sli}
+\Delta^{ip}\kappa_{k}^{0l}\epsilon_{plj}+\Delta^{sj}\kappa_{k}^{l0}\epsilon_{sli})\kappa_{q}^{ir}\epsilon_{jrn}\left\langle\Sigma_{0n}\right\rangle
\\\nonumber
&+&(\Delta^{s0}\kappa_{k}^{l0}+\Delta^{sp}\kappa_{k}^{lp})\kappa_{q}^{r0}\epsilon_{sli}\epsilon_{irn}\left\langle\Sigma_{n0}\right\rangle
+(\Delta^{0p}\kappa_{k}^{il}\epsilon_{plj}+\Delta^{s0}\kappa_{k}^{lj}\epsilon_{sli}
+\Delta^{ip}\kappa_{k}^{0l}\epsilon_{plj}+\Delta^{sj}\kappa_{k}^{l0}\epsilon_{sli})\kappa_{q}^{rj}\epsilon_{irn}\left\langle\Sigma_{n0}\right\rangle
\\\nonumber
&+&(\Delta^{0p}\kappa_{q}^{0l}+\Delta^{sp}\kappa_{q}^{sl})\kappa_{k}^{nr}\epsilon_{plj}\epsilon_{jrm}\left\langle\Sigma_{nm}\right\rangle
+(\Delta^{0p}\kappa_{q}^{nl}\epsilon_{plj}+\Delta^{s0}\kappa_{q}^{lj}\epsilon_{sln}
+\Delta^{np}\kappa_{q}^{0l}\epsilon_{plj}+\Delta^{sj}\kappa_{q}^{l0}\epsilon_{sln})\kappa_{k}^{0r}\epsilon_{jrm}\left\langle\Sigma_{nm}\right\rangle
\\\nonumber
&+&(\Delta^{s0}\kappa_{q}^{l0}+\Delta^{sp}\kappa_{q}^{lp})\kappa_{k}^{rm}\epsilon_{sli}\epsilon_{irn}\left\langle\Sigma_{nm}\right\rangle
+(\Delta^{0p}\kappa_{q}^{il}\epsilon_{plm}+\Delta^{s0}\kappa_{q}^{lm}\epsilon_{sli}
+\Delta^{ip}\kappa_{q}^{0l}\epsilon_{plm}+\Delta^{sm}\kappa_{q}^{l0}\epsilon_{sli})\kappa_{k}^{r0}\epsilon_{irn}\left\langle\Sigma_{nm}\right\rangle
\\\nonumber
&+&(\Delta^{0p}\kappa_{q}^{0l}+\Delta^{sp}\kappa_{q}^{sl})\kappa_{k}^{0r}\epsilon_{plj}\epsilon_{jrn}\left\langle\Sigma_{0n}\right\rangle
+(\Delta^{0p}\kappa_{q}^{il}\epsilon_{plj}+\Delta^{s0}\kappa_{q}^{lj}\epsilon_{sli}
+\Delta^{ip}\kappa_{q}^{0l}\epsilon_{plj}+\Delta^{sj}\kappa_{q}^{l0}\epsilon_{sli})\kappa_{q}^{ir}\epsilon_{jrn}\left\langle\Sigma_{0n}\right\rangle
\\\nonumber
&+&(\Delta^{s0}\kappa_{q}^{l0}+\Delta^{sp}\kappa_{q}^{lp})\kappa_{k}^{r0}\epsilon_{sli}\epsilon_{irn}\left\langle\Sigma_{n0}\right\rangle
+(\Delta^{0p}\kappa_{q}^{il}\epsilon_{plj}+\Delta^{s0}\kappa_{q}^{lj}\epsilon_{sli}
+\Delta^{ip}\kappa_{q}^{0l}\epsilon_{plj}+\Delta^{sj}\kappa_{q}^{l0}\epsilon_{sli})\kappa_{q}^{rj}\epsilon_{irn}\left\langle\Sigma_{n0}\right\rangle\Big]
\\\nonumber
&+&4\omega_{q}\Big[
(\Delta^{sp}\lambda_{k}^{sl}+\Delta^{0p}\lambda_{k}^{0l})\lambda_{q}^{nr}\epsilon_{plj}\epsilon_{jrm}\left\langle\Sigma_{nm}\right\rangle
+(\Delta^{0p}\lambda_{k}^{nl}\epsilon_{plj}+\Delta^{s0}\lambda_{k}^{lj}\epsilon_{sln}+\Delta^{np}\lambda_{k}^{0l}\epsilon_{plj}+\Delta^{sj}\lambda_{k}^{l0}\epsilon_{sln})\lambda_{q}^{0r}\epsilon_{jrm}\left\langle\Sigma_{nm}\right\rangle
\\\nonumber
&+&(\Delta^{s0}\lambda_{k}^{l0}+\Delta^{sp}\lambda_{k}^{lp})\lambda_{q}^{rm}\epsilon_{sli}\epsilon_{irn}\left\langle\Sigma_{nm}\right\rangle
+(\Delta^{0p}\lambda_{k}^{il}\epsilon_{plm}+\Delta^{s0}\lambda_{k}^{lm}\epsilon_{sli}+\Delta^{ip}\lambda_{k}^{0l}\epsilon_{plm}+\Delta^{sm}\lambda_{k}^{l0}\epsilon_{sli})\lambda_{q}^{r0}\epsilon_{irn}\left\langle\Sigma_{nm}\right\rangle
\\\nonumber
&+&(\Delta^{sp}\lambda_{k}^{sl}+\Delta^{0p}\lambda_{k}^{0l})\lambda_{q}^{0r}\epsilon_{plj}\epsilon_{jrn}\left\langle\Sigma_{0n}\right\rangle
+(\Delta^{0p}\lambda_{k}^{il}\epsilon_{plj}+\Delta^{s0}\lambda_{k}^{lj}\epsilon_{sli}+\Delta^{ip}\lambda_{k}^{0l}\epsilon_{plj}+\Delta^{sj}\lambda_{k}^{l0}\epsilon_{sli})\lambda_{q}^{ir}\epsilon_{jrn}\left\langle\Sigma_{0n}\right\rangle
\\\nonumber
&+&(\Delta^{s0}\lambda_{k}^{l0}+\Delta^{sp}\lambda_{k}^{lp})\lambda_{q}^{r0}\epsilon_{sli}\epsilon_{irn}\left\langle\Sigma_{n0}\right\rangle
+(\Delta^{0p}\lambda_{k}^{il}\epsilon_{plj}+\Delta^{s0}\lambda_{k}^{lj}\epsilon_{sli}+\Delta^{ip}\lambda_{k}^{0l}\epsilon_{plj}+\Delta^{sj}\lambda_{k}^{l0}\epsilon_{sli})\lambda_{q}^{rj}\epsilon_{irn}\left\langle\Sigma_{n0}\right\rangle
\\\nonumber
&+&(\Delta^{sp}\lambda_{q}^{sl}+\Delta^{0p}\lambda_{q}^{0l})\lambda_{k}^{nr}\epsilon_{plj}\epsilon_{jrm}\left\langle\Sigma_{nm}\right\rangle
+(\Delta^{0p}\lambda_{q}^{nl}\epsilon_{plj}+\Delta^{s0}\lambda_{q}^{lj}\epsilon_{sln}+\Delta^{np}\lambda_{q}^{0l}\epsilon_{plj}+\Delta^{sj}\lambda_{q}^{l0}\epsilon_{sln})\lambda_{k}^{0r}\epsilon_{jrm}\left\langle\Sigma_{nm}\right\rangle
\\\nonumber
&+&(\Delta^{s0}\lambda_{q}^{l0}+\Delta^{sp}\lambda_{q}^{lp})\lambda_{k}^{rm}\epsilon_{sli}\epsilon_{irn}\left\langle\Sigma_{nm}\right\rangle
+(\Delta^{0p}\lambda_{q}^{il}\epsilon_{plm}+\Delta^{s0}\lambda_{q}^{lm}\epsilon_{sli}+\Delta^{ip}\lambda_{q}^{0l}\epsilon_{plm}+\Delta^{sm}\lambda_{q}^{l0}\epsilon_{sli})\lambda_{k}^{r0}\epsilon_{irn}\left\langle\Sigma_{nm}\right\rangle
\\\nonumber
&+&(\Delta^{sp}\lambda_{q}^{sl}+\Delta^{0p}\lambda_{q}^{0l})\lambda_{k}^{0r}\epsilon_{plj}\epsilon_{jrn}\left\langle\Sigma_{0n}\right\rangle
+(\Delta^{0p}\lambda_{q}^{il}\epsilon_{plj}+\Delta^{s0}\lambda_{q}^{lj}\epsilon_{sli}+\Delta^{ip}\lambda_{q}^{0l}\epsilon_{plj}+\Delta^{sj}\lambda_{q}^{l0}\epsilon_{sli})\lambda_{k}^{ir}\epsilon_{jrn}\left\langle\Sigma_{0n}\right\rangle
\\\nonumber
&+&(\Delta^{s0}\lambda_{q}^{l0}+\Delta^{sp}\lambda_{q}^{lp})\lambda_{k}^{r0}\epsilon_{sli}\epsilon_{irn}\left\langle\Sigma_{n0}\right\rangle
+(\Delta^{0p}\lambda_{q}^{il}\epsilon_{plj}+\Delta^{s0}\lambda_{q}^{lj}\epsilon_{sli}+\Delta^{ip}\lambda_{q}^{0l}\epsilon_{plj}+\Delta^{sj}\lambda_{q}^{l0}\epsilon_{sli})\lambda_{k}^{rj}\epsilon_{irn}\left\langle\Sigma_{n0}\right\rangle\Big]
\Bigg\}\,.
\end{eqnarray}
The expectation values of \(\Sigma_{\alpha \beta}\) are defined by
\(
\left\langle\Sigma_{\alpha \beta}\right\rangle=
\mathrm{tr}\{\Sigma_{\alpha \beta}\exp(-\beta
H_{S B})\}/\mathrm{tr}\{\exp(-\beta H_{S B})\}\).
Here, \(H_{S B}\) is the Hamiltonian of the four-level system.

\section{The flow equations for the observables}\label{flow:sec:flowO}
\begin{eqnarray}\nonumber
\frac{dh^{0 0}}{dl}&=&4\sum\Big\{-\frac{1}{2}\Big(\lambda_{k}^{0 i}\mu_{k}^{0
i}+\lambda_{k}^{i 0}\mu_{k}^{i 0}+\lambda_{k}^{0 0}\mu_{k}^{0
0}+\lambda_{k}^{i j}\mu_{k}^{i j}
+\kappa_{k}^{0 i}\nu_{k}^{0
i}+\kappa_{k}^{i 0}\nu_{k}^{i 0}+\kappa_{k}^{0 0}\nu_{k}^{0
0}+\kappa_{k}^{i j}\nu_{k}^{i j}\Big)\omega_{k}
\\\nonumber
&-&(\Delta^{0p}\kappa_{k}^{0l}+\Delta^{sp}\kappa_{k}^{sl})\mu_{k}^{0j}\epsilon_{plj}
-(\Delta^{s0}\kappa_{k}^{l0}+\Delta^{sp}\kappa_{k}^{lp})\mu_{k}^{i0}\epsilon_{sli}
-(\Delta^{0p}\kappa_{k}^{il}\epsilon_{plj}+\Delta^{s0}\kappa_{k}^{lj}\epsilon_{sli}
+\Delta^{ip}\kappa_{k}^{0l}\epsilon_{plj}+\Delta^{sj}\kappa_{k}^{l0}\epsilon_{sli})\mu_{k}^{ij}
\\\nonumber
&+&(\Delta^{sp}\lambda_{k}^{sl}+\Delta^{0p}\lambda_{k}^{0l})\nu_{k}^{0j}\epsilon_{plj}
+(\Delta^{s0}\lambda_{k}^{l0}+\Delta^{sp}\lambda_{k}^{lp})\nu_{k}^{i0}\epsilon_{sli}
+(\Delta^{0p}\lambda_{k}^{il}\epsilon_{plj}+\Delta^{s0}\lambda_{k}^{lj}\epsilon_{sli}+\Delta^{ip}\lambda_{k}^{0l}\epsilon_{plj}+\Delta^{sj}\lambda_{k}^{l0}\epsilon_{sli})\nu_{k}^{ij}\Big\}
\\\nonumber
\frac{dh^{0 n}}{dl}&=&4\sum_{k}\Big\{\frac{1}{2}\Big(-\lambda_{k}^{i 0}\mu_{k}^{i
n}-\lambda_{k}^{0 0}\mu_{k}^{0 n}-\lambda_{k}^{i n}\mu_{k}^{i
0}-\lambda_{k}^{0 n}\mu_{k}^{0 0}
-\kappa_{k}^{i 0}\nu_{k}^{i
n}-\kappa_{k}^{0 0}\nu_{k}^{0 n}-\kappa_{k}^{i n}\nu_{k}^{i
0}-\kappa_{k}^{0 n}\nu_{k}^{0 0}
\\\nonumber
&+&[\kappa_{k}^{0 i}\mu_{k}^{0
l}+\kappa_{k}^{j i}\mu_{k}^{j l}-\lambda_{k}^{0 i}\nu_{k}^{0
l}-\lambda_{k}^{j i}\nu_{k}^{j l}]\epsilon_{i l n}(2
n_{k}+1)\Big)\omega_{k}
\\\nonumber
&-&\big[(\Delta^{sp}\lambda_{k}^{sl}
+\Delta^{0p}\lambda_{k}^{0l})\mu_{k}^{0r}\epsilon_{plj}\epsilon_{jrn}
+(\Delta^{0p}\lambda_{k}^{il}\epsilon_{plj}+\Delta^{s0}\lambda_{k}^{lj}\epsilon_{sli}+\Delta^{ip}\lambda_{k}^{0l}\epsilon_{plj}+\Delta^{sj}\lambda_{k}^{l0}\epsilon_{sli})\mu_{k}^{ir}\epsilon_{jrn}
\\\nonumber
&+&(\Delta^{0p}\kappa_{k}^{0l}+\Delta^{sp}\kappa_{k}^{sl})\nu_{k}^{0r}\epsilon_{plj}\epsilon_{jrn}
+(\Delta^{0p}\kappa_{k}^{il}\epsilon_{plj}+\Delta^{s0}\kappa_{k}^{lj}\epsilon_{sli}
+\Delta^{ip}\kappa_{k}^{0l}\epsilon_{plj}+\Delta^{sj}\kappa_{k}^{l0}\epsilon_{sli})\nu_{k}^{ir}\epsilon_{jrn}\big](2n_{k}+1)
\\\nonumber
&-&(\Delta^{0p}\kappa_{k}^{0l}+\Delta^{sp}\kappa_{k}^{sl})\mu_{k}^{00}\epsilon_{pln}
-(\Delta^{s0}\kappa_{k}^{l0}+\Delta^{sp}\kappa_{k}^{lp})\mu_{k}^{in}\epsilon_{sli}
-(\Delta^{0p}\kappa_{k}^{il}\epsilon_{pln}+\Delta^{s0}\kappa_{k}^{ln}\epsilon_{sli}
+\Delta^{ip}\kappa_{k}^{0l}\epsilon_{pln}+\Delta^{sn}\kappa_{k}^{l0}\epsilon_{sli})\mu_{k}^{i0}
\\\nonumber
&+&(\Delta^{sp}\lambda_{k}^{sl}+\Delta^{0p}\lambda_{k}^{0l})\nu_{k}^{00}\epsilon_{pln}
+(\Delta^{0p}\lambda_{k}^{il}\epsilon_{pln}+\Delta^{s0}\lambda_{k}^{ln}\epsilon_{sli}+\Delta^{ip}\lambda_{k}^{0l}\epsilon_{pln}+\Delta^{sn}\lambda_{k}^{l0}\epsilon_{sli})\nu_{k}^{i0}
+(\Delta^{s0}\lambda_{k}^{l0}+\Delta^{sp}\lambda_{k}^{lp})\nu_{k}^{in}\epsilon_{sli}\Big\}
\end{eqnarray}
\begin{eqnarray}\nonumber
\frac{dh^{n 0}}{dl}&=&4\sum_{k}\Big\{\frac{1}{2}\Big(-\lambda_{k}^{0 i}\mu_{k}^{n
i}-\lambda_{k}^{0 0}\mu_{k}^{n 0}-\lambda_{k}^{n i}\mu_{k}^{0
i}-\lambda_{k}^{n 0}\mu_{k}^{0 0}
-\kappa_{k}^{0 i}\nu_{k}^{n
i}-\kappa_{k}^{0 0}\nu_{k}^{n 0}-\kappa_{k}^{n i}\nu_{k}^{0
i}-\kappa_{k}^{n 0}\nu_{k}^{0 0}
\\\nonumber
&+&[\kappa_{k}^{i 0}\mu_{k}^{l
0}+\kappa_{k}^{i j}\mu_{k}^{l j}-\lambda_{k}^{i 0}\nu_{k}^{l
0}-\lambda_{k}^{i j}\nu_{k}^{l j}]\epsilon_{i l n}(2
n_{k}+1)\Big)\omega_{k}
\\\nonumber
&-&\big[(\Delta^{0p}\lambda_{k}^{il}\epsilon_{plj}+\Delta^{s0}\lambda_{k}^{lj}\epsilon_{sli}+\Delta^{ip}\lambda_{k}^{0l}\epsilon_{plj}+\Delta^{sj}\lambda_{k}^{l0}\epsilon_{sli})\mu_{k}^{rj}\epsilon_{irn}
+(\Delta^{s0}\lambda_{k}^{l0}\epsilon_{sli}
+\Delta^{sp}\lambda_{k}^{lp}\epsilon_{sli})\mu_{k}^{r0}\epsilon_{irn}
\\\nonumber
&+&(\Delta^{s0}\kappa_{k}^{l0}\epsilon_{sli}
+\Delta^{sp}\kappa_{k}^{lp}\epsilon_{sli})\nu_{k}^{r0}\epsilon_{irn}
+(\Delta^{0p}\kappa_{k}^{il}\epsilon_{plj}+\Delta^{s0}\kappa_{k}^{lj}\epsilon_{sli}
+\Delta^{ip}\kappa_{k}^{0l}\epsilon_{plj}+\Delta^{sj}\kappa_{k}^{l0}\epsilon_{sli})\nu_{k}^{rj}\epsilon_{irn}\big](2n_{k}+1)
\\\nonumber
&-&(\Delta^{0p}\kappa_{k}^{nl}\epsilon_{plj}+\Delta^{s0}\kappa_{k}^{lj}\epsilon_{sln}
+\Delta^{np}\kappa_{k}^{0l}\epsilon_{plj}+\Delta^{sj}\kappa_{k}^{l0}\epsilon_{sln})\mu_{k}^{0j}
-(\Delta^{s0}\kappa_{k}^{l0}+\Delta^{sp}\kappa_{k}^{lp})\mu_{k}^{00}\epsilon_{sln}
-(\Delta^{0p}\kappa_{k}^{0l}+\Delta^{sp}\kappa_{k}^{sl})\mu_{k}^{nj}\epsilon_{plj}
\\\nonumber
&+&(\Delta^{0p}\lambda_{k}^{nl}\epsilon_{plj}+\Delta^{s0}\lambda_{k}^{lj}\epsilon_{sln}+\Delta^{np}\lambda_{k}^{0l}\epsilon_{plj}+\Delta^{sj}\lambda_{k}^{l0}\epsilon_{sln})\nu_{k}^{0j}
+(\Delta^{sp}\lambda_{k}^{sl}+\Delta^{0p}\lambda_{k}^{0l})\nu_{k}^{nj}\epsilon_{plj}
+(\Delta^{s0}\lambda_{k}^{l0}+\Delta^{sp}\lambda_{k}^{lp})\nu_{k}^{00}\epsilon_{sln}\Big\}
\\\nonumber
\frac{dh^{n m}}{dl}&=&4\sum_{k}\Big\{\frac{1}{2}\Big(-\lambda_{k}^{0 0}\mu_{k}^{n m}-\lambda_{k}^{0 m}\mu_{k}^{n 0}-\lambda_{k}^{n 0}\mu_{k}^{0
m}-\lambda_{k}^{n m}\mu_{k}^{0 0}
-\kappa_{k}^{0 0}\nu_{k}^{n
m}-\kappa_{k}^{0 m}\nu_{k}^{n 0}
-\kappa_{k}^{n 0}\nu_{k}^{0
m}-\kappa_{k}^{n m}\nu_{k}^{0 0}
\\\nonumber
&+&(\lambda_{k}^{i j}\mu_{k}^{p
l}+\kappa_{k}^{i j}\nu_{k}^{p
l})\epsilon_{i p n}\epsilon_{j l m}
\\\nonumber
&+&\big[(\kappa_{k}^{i 0}\mu_{k}^{l m}+\kappa_{k}^{i m}\mu_{k}^{l
0}-\lambda_{k}^{i 0}\nu_{k}^{l m}-\lambda_{k}^{i m}\nu_{k}^{l
0})\epsilon_{i l n}
+(\kappa_{k}^{0 i}\mu_{k}^{n l}+\kappa_{k}^{n
i}\mu_{k}^{0 l}-\lambda_{k}^{0 i}\nu_{k}^{n l}-\lambda_{k}^{n
i}\nu_{k}^{0 l})\epsilon_{i l m}\big](2 n_{k}+1)\Big)\omega_{k}
\\\nonumber
&-&\big[(\Delta^{sp}\lambda_{k}^{sl}+\Delta^{0p}\lambda_{k}^{0l})\mu_{k}^{nr}\epsilon_{plj}\epsilon_{jrm}
+(\Delta^{0p}\lambda_{k}^{nl}\epsilon_{plj}+\Delta^{s0}\lambda_{k}^{lj}\epsilon_{sln}+\Delta^{np}\lambda_{k}^{0l}\epsilon_{plj}+\Delta^{sj}\lambda_{k}^{l0}\epsilon_{sln})\mu_{k}^{0r}\epsilon_{jrm}
\\\nonumber
&+&(\Delta^{s0}\lambda_{k}^{l0}+\Delta^{sp}\lambda_{k}^{lp})\mu_{k}^{rm}\epsilon_{sli}\epsilon_{irn}
+(\Delta^{0p}\lambda_{k}^{il}\epsilon_{plm}+\Delta^{s0}\lambda_{k}^{lm}\epsilon_{sli}+\Delta^{ip}\lambda_{k}^{0l}\epsilon_{plm}+\Delta^{sm}\lambda_{k}^{l0}\epsilon_{sli})\mu_{k}^{r0}\epsilon_{irn}
\\\nonumber
&+&(\Delta^{0p}\kappa_{k}^{0l}+\Delta^{sp}\kappa_{k}^{sl})\nu_{k}^{nr}\epsilon_{plj}\epsilon_{jrm}
+(\Delta^{0p}\kappa_{k}^{nl}\epsilon_{plj}+\Delta^{s0}\kappa_{k}^{lj}\epsilon_{sln}
+\Delta^{np}\kappa_{k}^{0l}\epsilon_{plj}+\Delta^{sj}\kappa_{k}^{l0}\epsilon_{sln})\nu_{k}^{0r}\epsilon_{jrm}\\\nonumber
&+&(\Delta^{s0}\kappa_{k}^{l0}+\Delta^{sp}\kappa_{k}^{lp})\nu_{k}^{rm}\epsilon_{sli}\epsilon_{irn}
+(\Delta^{0p}\kappa_{k}^{il}\epsilon_{plm}+\Delta^{s0}\kappa_{k}^{lm}\epsilon_{sli}
+\Delta^{ip}\kappa_{k}^{0l}\epsilon_{plm}+\Delta^{sm}\kappa_{k}^{l0}\epsilon_{sli})\nu_{k}^{r0}\epsilon_{irn}\big](2n_{k}+1)
\\\nonumber
&-&(\Delta^{0p}\kappa_{k}^{0l}+\Delta^{sp}\kappa_{k}^{sl})\mu_{k}^{n0}\epsilon_{plm}
+(\Delta^{0p}\kappa_{k}^{il}\epsilon_{plj}+\Delta^{s0}\kappa_{k}^{lj}\epsilon_{sli}
+\Delta^{ip}\kappa_{k}^{0l}\epsilon_{plj}+\Delta^{sj}\kappa_{k}^{l0}\epsilon_{sli})\mu_{k}^{tr}\epsilon_{itn}\epsilon_{jrm}
\\\nonumber
&-&(\Delta^{s0}\kappa_{k}^{l0}+\Delta^{sp}\kappa_{k}^{lp})\mu_{k}^{0m}\epsilon_{sln}
-(\Delta^{0p}\kappa_{k}^{nl}\epsilon_{plm}+\Delta^{s0}\kappa_{k}^{lm}\epsilon_{sln}
+\Delta^{np}\kappa_{k}^{0l}\epsilon_{plm}+\Delta^{sm}\kappa_{k}^{l0}\epsilon_{sln})\mu_{k}^{00}
\\\nonumber
&+&(\Delta^{sp}\lambda_{k}^{sl}+\Delta^{0p}\lambda_{k}^{0l})\nu_{k}^{n0}\epsilon_{plm}
-(\Delta^{0p}\lambda_{k}^{il}\epsilon_{plj}+\Delta^{s0}\lambda_{k}^{lj}\epsilon_{sli}+\Delta^{ip}\lambda_{k}^{0l}\epsilon_{plj}+\Delta^{sj}\lambda_{k}^{l0}\epsilon_{sli})\nu_{k}^{tr}\epsilon_{itn}\epsilon_{jrm}
\\\nonumber
&+&(\Delta^{s0}\lambda_{k}^{l0}+\Delta^{sp}\lambda_{k}^{lp})\nu_{k}^{0m}\epsilon_{sln}
+(\Delta^{0p}\lambda_{k}^{nl}\epsilon_{plm}+\Delta^{s0}\lambda_{k}^{lm}\epsilon_{sln}+\Delta^{np}\lambda_{k}^{0l}\epsilon_{plm}+\Delta^{sm}\lambda_{k}^{l0}\epsilon_{sln})\nu_{k}^{00}\Big\}
\end{eqnarray}

\begin{eqnarray}
\frac{d\mu_{k}^{0 0}}{dl}&=&2\sum_{q}\eta_{k q}\mu_{q}^{0 0}
\\\nonumber
\frac{d\mu_{k}^{0 n}}{dl}&=&2(\kappa_{k}^{0 i}h^{0 l}+\kappa_{k}^{j
i}h^{j l})\epsilon_{i l n}\omega_{k}+2\sum_{q}\eta_{k q}\mu_{q}^{0 n}
\\\nonumber
&-&4\big[(\Delta^{sp}\lambda_{k}^{sl}+\Delta^{0p}\lambda_{k}^{0l})h^{0r}\epsilon_{plj}\epsilon_{jrn}
+(\Delta^{0p}\lambda_{k}^{il}\epsilon_{plj}+\Delta^{s0}\lambda_{k}^{lj}\epsilon_{sli}+\Delta^{ip}\lambda_{k}^{0l}\epsilon_{plj}+\Delta^{sj}\lambda_{k}^{l0}\epsilon_{sli})h^{ir}\epsilon_{jrn}\big]
\\\nonumber
\frac{d\mu_{k}^{n 0}}{dl}&=&2(\kappa_{k}^{i 0}h^{l
0}+\kappa_{k}^{i j}h^{l j})\epsilon_{i l n}\omega_{k}+2\sum_{q}\eta_{k
q}\mu_{q}^{n 0}
\\\nonumber
&-&4\big[(\Delta^{0p}\lambda_{k}^{il}\epsilon_{plj}+\Delta^{s0}\lambda_{k}^{lj}\epsilon_{sli}+\Delta^{ip}\lambda_{k}^{0l}\epsilon_{plj}+\Delta^{sj}\lambda_{k}^{l0}\epsilon_{sli})h^{rj}\epsilon_{irn}
+(\Delta^{s0}\lambda_{k}^{l0}+\Delta^{sp}\lambda_{k}^{lp})h^{r0}\epsilon_{sli}\epsilon_{irn}\big]
\\\nonumber
\frac{d\mu_{k}^{n m}}{dl}&=&2\Big((\kappa_{k}^{0 i}h^{n l}+\kappa_{k}^{n i}h^{0 l})\epsilon_{i
l m}+(\kappa_{k}^{i 0}h^{l m}+\kappa_{k}^{i m}h^{l
0})\epsilon_{i l n}\Big)\omega_{k}+2\sum_{q}\eta_{k q}\mu_{q}^{n m}
\\\nonumber
&-&4\big[
(\Delta^{sp}\lambda_{k}^{sl}+\Delta^{0p}\lambda_{k}^{0l})h^{nr}\epsilon_{plj}\epsilon_{jrm}
+(\Delta^{0p}\lambda_{k}^{nl}\epsilon_{plj}+\Delta^{s0}\lambda_{k}^{lj}\epsilon_{sln}+\Delta^{np}\lambda_{k}^{0l}\epsilon_{plj}+\Delta^{sj}\lambda_{k}^{l0}\epsilon_{sln})h^{0r}\epsilon_{jrm}
\\\nonumber
&+&(\Delta^{s0}\lambda_{k}^{l0}
+\Delta^{sp}\lambda_{k}^{lp})h^{rm}\epsilon_{sli}\epsilon_{irn}
+(\Delta^{0p}\lambda_{k}^{il}\epsilon_{plm}+\Delta^{s0}\lambda_{k}^{lm}\epsilon_{sli}+\Delta^{ip}\lambda_{k}^{0l}\epsilon_{plm}+\Delta^{sm}\lambda_{k}^{l0}\epsilon_{sli})h^{r0}\epsilon_{irn}
\big]
\end{eqnarray}
\begin{eqnarray}\nonumber
\frac{d\nu_{k}^{0 0}}{dl}&=&-2\sum_{q}\eta_{q k}\nu_{q}^{0 0}
\\\nonumber
\frac{d\nu_{k}^{0 n}}{dl}&=&-2(\lambda_{k}^{0 i}h^{0 l}+\lambda_{k}^{j
i}h^{j l})\epsilon_{i l n}\omega_{k}-2\sum_{q}\eta_{q k}\nu_{q}^{0 n}
\\\nonumber
&-&4\big[(\Delta^{0p}\kappa_{k}^{0l}+\Delta^{sp}\kappa_{k}^{sl})h^{0r}\epsilon_{plj}\epsilon_{jrn}
+(\Delta^{0p}\kappa_{k}^{il}\epsilon_{plj}+\Delta^{s0}\kappa_{k}^{lj}\epsilon_{sli}
+\Delta^{ip}\kappa_{k}^{0l}\epsilon_{plj}+\Delta^{sj}\kappa_{k}^{l0}\epsilon_{sli})h^{ir}\epsilon_{jrn}\big]
\\\nonumber
\frac{d\nu_{k}^{n 0}}{dl}&=&-2(\lambda_{k}^{i 0}h^{l 0}+\lambda_{k}^{i j}h^{l j})\epsilon_{i l n}\omega_{k}-2\sum_{q}\eta_{q k}\nu_{q}^{n
0}
\\\nonumber
&-&4\big[(\Delta^{0p}\kappa_{k}^{il}\epsilon_{plj}+\Delta^{s0}\kappa_{k}^{lj}\epsilon_{sli}
+\Delta^{ip}\kappa_{k}^{0l}\epsilon_{plj}+\Delta^{sj}\kappa_{k}^{l0}\epsilon_{sli})h^{rj}\epsilon_{irn}
+(\Delta^{s0}\kappa_{k}^{l0}+\Delta^{sp}\kappa_{k}^{lp})h^{r0}\epsilon_{sli}\epsilon_{irn}\big]
\\\nonumber
\frac{d\nu_{k}^{n m}}{dl}&=&-2\Big((\lambda_{k}^{0
i}h^{n l}+\lambda_{k}^{n i}h^{0 l})\epsilon_{i l
m}+(\lambda_{k}^{i 0}h^{l m}+\lambda_{k}^{i m}h^{l
0})\epsilon_{i l n}\Big)\omega_{k}-2\sum_{q}\eta_{q k}\nu_{q}^{n m}
\\\nonumber
&-&4\big[
(\Delta^{0p}\kappa_{k}^{0l}+\Delta^{sp}\kappa_{k}^{sl})h^{nr}\epsilon_{plj}\epsilon_{jrm}
+(\Delta^{0p}\kappa_{k}^{nl}\epsilon_{plj}+\Delta^{s0}\kappa_{k}^{lj}\epsilon_{sln}
+\Delta^{np}\kappa_{k}^{0l}\epsilon_{plj}+\Delta^{sj}\kappa_{k}^{l0}\epsilon_{sln})h^{0r}\epsilon_{jrm}
\\\nonumber
&+&(\Delta^{s0}\kappa_{k}^{l0}+\Delta^{sp}\kappa_{k}^{lp})h^{rm}\epsilon_{sli}\epsilon_{irn}
+(\Delta^{0p}\kappa_{k}^{il}\epsilon_{plm}+\Delta^{s0}\kappa_{k}^{lm}\epsilon_{sli}
+\Delta^{ip}\kappa_{k}^{0l}\epsilon_{plm}+\Delta^{sm}\kappa_{k}^{l0}\epsilon_{sli})h^{r0}\epsilon_{irn}
\big]\,.
\end{eqnarray}

\section{A special case, the biased system}\label{flow:sec:flowbias}
Here we present the flow equations for the specific case
\(\epsilon_{S}=0\), which was mainly used in this work. 
\begin{eqnarray}\nonumber
\frac{d\Delta^{0 0}}{dl}&=&\sum_{k}[-2\Big(\lambda_{k}^{3 1}\lambda_{k}^{31}+\lambda_{k}^{0 3}\lambda_{k}^{0
3}+\kappa_{k}^{21}\kappa_{k}^{21}+\kappa_{k}^{0 2}\kappa_{k}^{0
2}\Big)\omega_{k}
\\\nonumber
&+&8(-\Delta^{0 1}\lambda_{k}^{03}\kappa_{k}^{02}+\Delta^{33}\lambda_{k}^{31}\kappa_{k}^{02}-\Delta^{10}\lambda_{k}^{31}\kappa_{k}^{21}+\Delta^{22}\lambda_{k}^{03}\kappa_{k}^{21})]
\\\nonumber
\frac{d\Delta^{01}}{dl}&=&\sum_{k}4(\kappa_{k}^{02}\lambda_{k}^{03}\omega_{k}+\Delta^{01}\lambda_{k}^{03}\lambda_{k}^{03}-\Delta^{33}\lambda_{k}^{31}\lambda_{k}^{03}+\Delta^{01}\kappa_{k}^{02}\kappa_{k}^{02}-\Delta^{22}\kappa_{k}^{21}\kappa_{k}^{02})
\\\nonumber
\frac{d\Delta^{10}}{dl}&=&\sum_{k}4(\kappa_{k}^{21}\lambda_{k}^{31}\omega_{k}+\Delta^{10}\lambda_{k}^{31}\lambda_{k}^{31}-\Delta^{22}\lambda_{k}^{03}\lambda_{k}^{31}+\Delta^{10}\kappa_{k}^{21}\kappa_{k}^{21}-\Delta^{33}\kappa_{k}^{02}\kappa_{k}^{21})
\\\nonumber
\frac{d\Delta^{22}}{dl}&=&\sum_{k}4(-\kappa_{k}^{21}\lambda_{k}^{03}\omega_{k}-\Delta^{10}\lambda_{k}^{31}\lambda_{k}^{03}+\Delta^{22}\lambda_{k}^{03}\lambda_{k}^{03}-\Delta^{01}\kappa_{k}^{02}\kappa_{k}^{21}+\Delta^{22}\kappa_{k}^{21}\kappa_{k}^{21})
\\\nonumber
\frac{d\Delta^{33}}{dl}&=&\sum_{k}4(-\kappa_{k}^{02}\lambda_{k}^{31}\omega_{k}-\Delta^{01}\lambda_{k}^{31}\lambda_{k}^{03}+\Delta^{33}\lambda_{k}^{31}\lambda_{k}^{31}-\Delta^{10}\kappa_{k}^{21}\kappa_{k}^{02}+\Delta^{33}\kappa_{k}^{02}\kappa_{k}^{02})
\\\nonumber
\frac{d\lambda_{k}^{03}}{dl}&=&-\omega_{k}^2\lambda_{k}^{03}-4\kappa_{k}^{02}\Delta^{01}\omega_{k}+4\kappa_{k}^{21}\Delta^{22}\omega_{k}+2\sum_{q}\eta_{kq}\lambda_{q}^{03}
\\\nonumber
&-&4\Delta^{01}\Delta^{01}\lambda_{k}^{03}+4\Delta^{33}\Delta^{01}\lambda_{k}^{31}+4\Delta^{10}\Delta^{22}\lambda_{k}^{31}-4\Delta^{22}\Delta^{22}\lambda_{k}^{03}
\\\nonumber
\frac{d\lambda_{k}^{31}}{dl}&=&-\omega_{k}^2\lambda_{k}^{31}-4\kappa_{k}^{21}\Delta^{10}\omega_{k}+4\kappa_{k}^{02}\Delta^{33}\omega_{k}+2\sum_{q}\eta_{kq}\lambda_{q}^{31}
\\\nonumber
&+&4\Delta^{01}\Delta^{33}\lambda_{k}^{03}-4\Delta^{33}\Delta^{33}\lambda_{k}^{31}-4\Delta^{10}\Delta^{10}\lambda_{k}^{31}+4\Delta^{22}\Delta^{10}\lambda_{k}^{03}
\end{eqnarray}
\begin{eqnarray}\nonumber
\frac{d\kappa_{k}^{02}}{dl}&=&-\omega_{k}^2\kappa_{k}^{02}-4\lambda_{k}^{03}\Delta^{01}\omega_{k}+4\lambda_{k}^{31}\Delta^{33}\omega_{k}-2\sum_{q}\eta_{qk}\kappa_{q}^{02}
\\\nonumber
&-&4\Delta^{01}\Delta^{01}\kappa_{k}^{02}+4\Delta^{22}\Delta^{01}\kappa_{k}^{21}+4\Delta^{10}\Delta^{33}\kappa_{k}^{21}-4\Delta^{33}\Delta^{33}\kappa_{k}^{02}
\\\nonumber
\frac{d\kappa_{k}^{21}}{dl}&=&-\omega_{k}^2\kappa_{k}^{21}-4\lambda_{k}^{31}\Delta^{10}\omega_{k}+4\lambda_{k}^{03}\Delta^{22}\omega_{k}-2\sum_{q}\eta_{qk}\kappa_{q}^{21}
\\\nonumber
&+&4\Delta^{01}\Delta^{22}\kappa_{k}^{02}-4\Delta^{22}\Delta^{22}\kappa_{k}^{21}-4\Delta^{10}\Delta^{10}\kappa_{k}^{21}+4\Delta^{33}\Delta^{10}\kappa_{k}^{02}
\\\nonumber
\frac{d\mu_{k}^{11}}{dl}&=&2\kappa_{k}^{02}h^{13}\omega_{k}+2\kappa_{k}^{21}h^{30}\omega_{k}+2\sum_{q}\eta_{kq}\mu_{q}^{11}
\\\nonumber
&+&4\Delta^{01}\lambda_{k}^{03}h^{13}-4\Delta^{33}\lambda_{k}^{31}h^{13}+4\Delta^{10}\lambda_{k}^{31}h^{30}-4\Delta^{22}\lambda_{k}^{03}h^{30}
\\\nonumber
\frac{dh^{13}}{dl}&=&\sum_{k}(-2\kappa_{k}^{02}\mu_{k}^{11}\omega_{k}
-4\Delta^{01}\lambda_{k}^{03}\mu_{k}^{11}+4\Delta^{33}\lambda_{k}^{31}\mu_{k}^{11})
\\
\frac{dh^{30}}{dl}&=&\sum_{k}(-2\kappa_{k}^{21}\mu_{k}^{11}\omega_{k}
-4\Delta^{10}\lambda_{k}^{31}\mu_{k}^{11}+4\Delta^{22}\lambda_{k}^{03}\mu_{k}^{11})\,.
\end{eqnarray}
The first equation for the coefficient \(\Delta^{00}\) constitutes an
energy renormalization. It is not necessary to take this equation into account for the numerical integration of the
correlation functions.
The \(\eta_{kq}\) coefficients are given by (\(k\neq q\))
\begin{eqnarray}\nonumber
\eta_{k q}&=&\frac{1}{\omega_{q}^2-\omega_{k}^2}\Bigg\{2\omega_{k}\Big[\big(\lambda_{k}^{0 3}\kappa_{q}^{21}\left\langle\Sigma_{22}\right\rangle+\lambda_{k}^{31}\kappa_{q}^{0
2}\left\langle\Sigma_{33}\right\rangle
-\lambda_{k}^{0 3}\kappa_{q}^{0
2}\left\langle\Sigma_{0 1}\right\rangle
-\lambda_{k}^{31}\kappa_{q}^{21}\left\langle\Sigma_{1 0}\right\rangle\big)\omega_{k}
\\
&+&\big(
\lambda_{q}^{0 3}\kappa_{k}^{21}\left\langle\Sigma_{22}\right\rangle
+\lambda_{q}^{31}\kappa_{k}^{0
2}\left\langle\Sigma_{33}\right\rangle
-\lambda_{q}^{0 3}\kappa_{k}^{0
2}\left\langle\Sigma_{0 1}\right\rangle
-\lambda_{q}^{31}\kappa_{k}^{21}\left\langle\Sigma_{10}\right\rangle\big)\omega_{q}\Big]
\\\nonumber
&-&2\omega_{q}\Big[\big(-\kappa_{k}^{0 2}\lambda_{q}^{3
1}\left\langle\Sigma_{3 3}\right\rangle
-\kappa_{k}^{2 1}\lambda_{q}^{0
3}\left\langle\Sigma_{2 2}\right\rangle
+\kappa_{k}^{0 2}\lambda_{q}^{0
3}\left\langle\Sigma_{0 1}\right\rangle
+\kappa_{k}^{2 1}\lambda_{q}^{3
1}\left\langle\Sigma_{1 0}\right\rangle\big)\omega_{k}
\\\nonumber
&+&\big(-\kappa_{q}^{0 2}\lambda_{k}^{3
1}\left\langle\Sigma_{3 3}\right\rangle
-\kappa_{q}^{2 1}\lambda_{k}^{0
3}\left\langle\Sigma_{2 2}\right\rangle
+\kappa_{q}^{0 2}\lambda_{k}^{0
3}\left\langle\Sigma_{0 1}\right\rangle
+\kappa_{q}^{2 1}\lambda_{k}^{3
1}\left\langle\Sigma_{1 0}\right\rangle\big)\omega_{q}\Big]
\\\nonumber
&+&4\omega_{k}\big(-\Delta^{01}\kappa_{k}^{0 2}\kappa_{q}^{0 2}\left\langle\Sigma_{01}\right\rangle
+\Delta^{01}\kappa_{k}^{0
2}\kappa_{q}^{21}\left\langle\Sigma_{22}\right\rangle
+\Delta^{10}\kappa_{k}^{21}\kappa_{q}^{02}\left\langle\Sigma_{33}\right\rangle
-\Delta^{10}\kappa_{k}^{21}\kappa_{q}^{21}\left\langle\Sigma_{10}\right\rangle
\\\nonumber
&+&\Delta^{22}\kappa_{k}^{21}\kappa_{q}^{0 2}\left\langle\Sigma_{01}\right\rangle
-\Delta^{22}\kappa_{k}^{21}\kappa_{q}^{21}\left\langle\Sigma_{22}\right\rangle
-\Delta^{33}\kappa_{k}^{02}\kappa_{q}^{02}\left\langle\Sigma_{33}\right\rangle
+\Delta^{33}\kappa_{k}^{02}\kappa_{q}^{21}\left\langle\Sigma_{10}\right\rangle
\big)
\\\nonumber
&+&4\omega_{k}\big(-\Delta^{01}\kappa_{q}^{0 2}\kappa_{k}^{0 2}\left\langle\Sigma_{01}\right\rangle
+\Delta^{01}\kappa_{q}^{0
2}\kappa_{k}^{21}\left\langle\Sigma_{22}\right\rangle
+\Delta^{10}\kappa_{q}^{21}\kappa_{k}^{02}\left\langle\Sigma_{33}\right\rangle
-\Delta^{10}\kappa_{q}^{21}\kappa_{k}^{21}\left\langle\Sigma_{10}\right\rangle
\\\nonumber
&+&\Delta^{22}\kappa_{q}^{21}\kappa_{k}^{0 2}\left\langle\Sigma_{01}\right\rangle
-\Delta^{22}\kappa_{q}^{21}\kappa_{k}^{21}\left\langle\Sigma_{22}\right\rangle
-\Delta^{33}\kappa_{q}^{02}\kappa_{k}^{02}\left\langle\Sigma_{33}\right\rangle
+\Delta^{33}\kappa_{q}^{02}\kappa_{k}^{21}\left\langle\Sigma_{10}\right\rangle
\big)
\\\nonumber
&+&4\omega_{q}\big(-\Delta^{01}\lambda_{k}^{0 3}\lambda_{q}^{0 3}\left\langle\Sigma_{01}\right\rangle
+\Delta^{01}\lambda_{k}^{03}\lambda_{q}^{31}\left\langle\Sigma_{33}\right\rangle
+\Delta^{10}\lambda_{k}^{31}\lambda_{q}^{03}\left\langle\Sigma_{22}\right\rangle
-\Delta^{10}\lambda_{k}^{31}\lambda_{q}^{31}\left\langle\Sigma_{10}\right\rangle
\\\nonumber
&-&\Delta^{22}\lambda_{k}^{03}\lambda_{q}^{0 3}\left\langle\Sigma_{22}\right\rangle
+\Delta^{22}\lambda_{k}^{03}\lambda_{q}^{31}\left\langle\Sigma_{10}\right\rangle
+\Delta^{33}\lambda_{k}^{31}\lambda_{q}^{03}\left\langle\Sigma_{01}\right\rangle
-\Delta^{33}\lambda_{k}^{31}\lambda_{q}^{31}\left\langle\Sigma_{33}\right\rangle
\big)
\\\nonumber
&+&4\omega_{q}\big(-\Delta^{01}\lambda_{q}^{0 3}\lambda_{k}^{0 3}\left\langle\Sigma_{01}\right\rangle
+\Delta^{01}\lambda_{q}^{03}\lambda_{k}^{31}\left\langle\Sigma_{33}\right\rangle
+\Delta^{10}\lambda_{q}^{31}\lambda_{k}^{03}\left\langle\Sigma_{22}\right\rangle
-\Delta^{10}\lambda_{q}^{31}\lambda_{k}^{31}\left\langle\Sigma_{10}\right\rangle
\\\nonumber
&-&\Delta^{22}\lambda_{q}^{03}\lambda_{k}^{0 3}\left\langle\Sigma_{22}\right\rangle
+\Delta^{22}\lambda_{q}^{03}\lambda_{k}^{31}\left\langle\Sigma_{10}\right\rangle
+\Delta^{33}\lambda_{q}^{31}\lambda_{k}^{03}\left\langle\Sigma_{01}\right\rangle
-\Delta^{33}\lambda_{q}^{31}\lambda_{k}^{31}\left\langle\Sigma_{33}\right\rangle
\big)
\Bigg\}\,.
\end{eqnarray}
For \(\Delta>0\), and \(\Delta_{S}>0\) 
the only non-vanishing expectation values \(\langle\Sigma_{\alpha \beta}\rangle\) are
\begin{eqnarray}\nonumber
\left\langle\Sigma_{0 0}\right\rangle&=&\left\langle\Sigma_{1 1}\right\rangle=1
\\\nonumber
\left\langle\Sigma_{0
1}\right\rangle&=&\left\langle\Sigma_{1 0}\right\rangle=
-\frac{\Delta+\Delta_{S}}{\sqrt{g^2+(\Delta+\Delta_{S})^2}}
\\
\left\langle\Sigma_{2 2}\right\rangle&=&
-\left\langle\Sigma_{3 3}\right\rangle=\frac{g}
{\sqrt{g^2+(\Delta+\Delta_{S})^2}}\,.
\end{eqnarray}
\end{widetext}

\end{document}